\documentclass{aa}
\usepackage{graphicx}
\newcommand{\gsim}{\,\raisebox{0.2em}{$>$}\!\!\!\!\!
\raisebox{-0.25em}{$\sim$}\,}
\newcommand{\lsim}{\,\raisebox{0.2em}{$<$}\!\!\!\!\!
\raisebox{-0.25em}{$\sim$}\,}

\begin{document}
\title{ The theory of synchrotron emission from supernova remnants}

   \subtitle{}

\author{E.G. Berezhko
        \inst{1}
         \and
         H.J. V\"olk
         \inst{2}}

   \offprints{H.J.V\"olk}

   \institute{Institute of Cosmophysical Research and Aeronomy,
                     31 Lenin Ave., 677891 Yakutsk, Russia\\
              \email{berezhko@ikfia.ysn.ru}
         \and
             Max Planck Institut f\"ur Kernphysik,
                Postfach 103980, D-69029 Heidelberg, Germany\\
             \email{Heinrich.Voelk@mpi-hd.mpg.de}
             }

   \date{Received month day, year; accepted month day, year}

     \abstract{The time-dependent nonlinear kinetic theory for cosmic ray 
       (CR) acceleration 
       in supernova remnants (SNRs) is applied
       studying  the
       properties of the synchrotron emission from SNRs, in 
       particular,
       the surface brightness-diameter ($\Sigma-D$) relation. Detailed 
       numerical calculations are performed for the expected range of
       the relevant physical parameters, namely the ambient 
       density and the supernova explosion energy. The magnetic field
       in SNRs is assumed to be significantly amplified by the
       efficiently accelerating nuclear CR component. Due to the growing 
       number of
       accelerated CRs the expected SNR luminosity increases during the 
       free
       expansion phase, reaches a peak value at the 
       beginning of the Sedov phase and then decreases again, since in 
       this stage the overall CR number remains nearly constant,
       whereas the effective magnetic field diminishes with time.  
       The theoretically predicted brightness-diameter relation
       in the radio range in the Sedov phase is close to
       $\Sigma_\mathrm{R}\propto D^{-17/4}$. It 
       fits the observational data in a very 
       satisfactory way. The observed spread of $\Sigma_\mathrm{R}$ at a given
       SNR size $D$ is the result of the spread of supernova explosion 
       energies and interstellar medium densities.
       
}
\titlerunning{Theory of synchrotron emission from SNRs}

   \maketitle
%
%________________________________________________________________

\section{Introduction}
Supernova remnants (SNRs) are the main sources of energy for the
Interstellar Medium (ISM). They also
control the physical state of the
ISM which presumably includes the nonthermal component of Interstellar
matter, often called the Galactic Cosmic Rays (CRs). The synchrotron
emission of relativistic electrons plays an important role in the general
study of SNR properties and in CR production inside SNRs in particular.
All known SNRs are sources of radio-synchrotron emission. Several
Galactic SNRs were recently detected as sources of nonthermal X-ray
emission which is presumably also of synchrotron origin (e.g. Petre et
al. 2001).

The determination of the distances to the Galactic SNRs is an important
task which is often based on radio observations. When
there is no direct distance determination, estimates can be made by using
the radio surface brightness-to-diameter relationship $(\Sigma_\mathrm{R}-D)$.  It
is however not clear whether any functional correlation between
$\Sigma_\mathrm{R}(t)$ and $D(t)$ exists for individual objects during their
evolution in time $t$, and if so, for which physical reason (Green 1984;
Case \& Bhattacharya 1998).

CRs are widely accepted to be produced in SNRs by the diffusive shock
acceleration process at the outer blast wave (see e.g. Drury 1983;
Blandford \& Eichler 1987; Berezhko \& Krymsky 1988, for reviews).
The nonlinear kinetic theory of diffusive CR acceleration in SNRs (Berezhko et
al. 1996; Berezhko \& V\"olk 1997) is able to make quite definite
predictions about the energy spectrum and the spatial distribution of CR
nuclei and electrons at any given evolutionary epoch $t$, and about the
properties of the nonthermal radiation produced in SNRs due to these
accelerated CRs. Application of this theory to individual SNRs (Berezhko
et al. 2002, 2003a,b; V\"olk et al. 2002) has demonstrated its power in
explaining the observed SNR properties and in predicting 
new effects like magnetic field amplification, leading to the concentration of
the highest-energy electrons in a very thin shell just behind the
shock. At the same time there does not yet exist a systematic
study of the synchrotron emission expected during the different
evolutionary epochs of SNRs, possibly leading in particular to a
$\Sigma_\mathrm{R}-D$ relation. An interesting question is specifically whether or
not the observed $\Sigma_\mathrm{R}-D$ relation corresponds to the evolutionary
track of a single typical SNR, as was for example argued by Duric \&
Seaquist (1986). Such a study is the aim of the present paper.

Previous considerations (e.g. Reynolds \& Chevalier 1981; Duric \&
Seaquist 1986; Huang et al. 1994) show that to explain on average the
observed radio synchrotron emission one needs magnetic fields inside SNRs
that are much higher than those corresponding to typical ISM values,
$B_\mathrm{ISM}\approx 5$~$\mu$G. The Rayleigh-Taylor instability of the
contact discontinuity, separating ejecta and swept-up ISM, has been
considered as a possible mechanism of magnetic field amplification in SNRs
(Gull 1973; Fedorenko 1983; Duric \& Seaquist 1986). However, it is
unlikely that electrons which are accelerated at the outer blast wave and
produce radio emission can penetrate so deeply into the interior where the
magnetic field is amplified due to this instability. The typical energies
of such radio electrons is so small that their diffusion coefficient is
several orders of magnitude less than $R_\mathrm{s}V_\mathrm{s}$, the
product of shock radius and shock speed. This means that these electrons
are strongly tied to the downstream medium and that their diffusion is not
important. To play a role in the synchrotron emissivity of SNRs, magnetic
field amplification should occur in the same region where also the
energetic electrons are produced, that is at the shock front. Assuming
such a scenario Reynolds \& Chevalier (1981) concluded that several
percent of the shock energy should be converted into (amplified) magnetic
field energy to explain the observed properties of the radio emission
during the Sedov phase. If the magnetic field was not amplified, the
expected time dependence of the SNR synchrotron emissivity did not fit the
observations.

On the other hand, Luceck \& Bell (2000) came to the conclusion that a
strong magnetic field amplification near the shock can indeed be produced
nonlinearly by a very efficiently accelerated nuclear CR component. In
this case the CR energy content and streaming anisotropy along the
field is so high that magnetohydrodynamic waves can be strongly excited,
amplifying the magnetic field already upstream of
the shock, where the CR distribution is characterized by a large spatial
gradient. In addition, detailed observational
evidence has been obtained regarding the spatial fine structure of
nonthermal X-rays in young SNRs (Hwang et al. 2002; Vink \& Laming 2003;
Long et al. 2003; Bamba et al. 2003), which can be explained only if the
magnetic field at the outer shock and inside SNRs is indeed strongly
amplified (Berezhko et al. 2003a; Berezhko \& V\"olk 2004).

In the following we shall present a detailed study of the evolutionary
properties of the synchrotron emission of SNRs. There are a number of
distinct differences between the present and previous studies: (i) our
approach is based on the nonlinear kinetic theory self-consistently taking
into account the backreaction of the accelerating CRs on the SN shock
structure and remnant dynamics (ii) to reproduce the evolution of the
synchrotron emission we calculate self-consistent electron spectra which
take the synchrotron losses into account. This is done for every SNR
evolutionary stage starting from the very early time after the SN
explosion up to the late Sedov phase when the SN shock already becomes an
inefficient accelerator (iii) we take into account magnetic field
amplification near the outer SN shock due to CR backreaction on the
thermal plasma and its magnetic field (iv) to explain the spread of the
observed properties of the SNR synchrotron emission we perform our
calculations for different values of the relevant physical parameters such
as SN explosion energy and ISM density (v) since recently X-ray
synchrotron emission was detected for a number of SNRs, we study here the
expected properties of the SNR synchrotron emission for the entire
wavelength range, from the radio to the X-ray band.

\section{Model}

A supernova explosion ejects an expanding shell of matter with total
energy $E_\mathrm{sn}$ and mass $M_\mathrm{ej}$ into the surrounding ISM.
During an initial period the shell material has a broad distribution in
velocity $v$. The fastest part of these ejecta is described by a power law
\begin{equation}
dM_\mathrm{ej}/dv\propto v^\mathrm{2-k}
\label{Mej}
\end{equation}
with $k=7$~to~12 (e.g. Jones et al. 1981; Chevalier 1982). The interaction
of the ejecta with the ISM creates a strong shock which accelerates
particles. We note that more recently also exponential profiles for the
velocity distribution of the ejecta have been discussed (Dwarkadas \&
Chevalier 1998). However we believe that for the global synchrotron
emission the differences between these different approximations are not
significant. Thus we use here the power-law ansatz above.

Our nonlinear model (Berezhko et al. 1996; Berezhko \& V\"olk 1997)  is
based on a fully time-dependent solution of the CR transport equation
together with the gas dynamic equations in spherical symmetry. Since all
relevant equations, initial and boundary conditions for this model have
already been described in detail elsewhere (Berezhko et al. 1996;  
Berezhko \& V\"olk 1997; Berezhko \& Ksenofontov 1999), we do not present
them here and only briefly discuss the most important aspects below.

Due to the streaming instability CRs efficiently excite large-amplitude
magnetic fluctuations upstream of the SN shock (e.g.  
Bell 1978; Blandford \& Ostriker 1978; McKenzie \& V\"olk 1982). Since
these fluctuations scatter CRs extremely strongly, the
CR diffusion coefficient is assumed to be as small as the Bohm limit
\begin{equation}
\kappa (p)=\kappa (mc)(p/mc), 
\label{kappa}
\end{equation}
where $\kappa(mc)=mc^2/(3eB)$, $e$ and $m$ are the particle charge and mass,
$p$ denotes the particle momentum, $B$ is the magnetic field strength, 
and $c$ is the speed of light.

If $B$ is the pre-existing field in the surrounding ISM, then the Bohm
limit implies that the instability growth is restricted by some nonlinear
mechanism to the level $\delta B\sim B$, where $\delta B$ is the wave
field. The attempt to give a nonlinear description of the magnetic field
evolution in a numerical model ( Luceck \& Bell 2000) concluded that a
considerable field amplification should occur, resulting in what we call
the effective magnetic field.  Broadly speaking, it was expected that a
significant fraction of the shock ram pressure $\rho_\mathrm{0}
V_\mathrm{s}^2$ ($\rho_\mathrm{0}$ is the ISM density, $V_\mathrm{s}$ is
the shock speed) would be converted into magnetic field energy (Bell \&
Luceck 2001).

At the same time, from an analysis of the synchrotron {\it spectrum} of
SN~1006 (Berezhko et al. \cite{bkv02}), Cassiopeia~A (Berezhko et al.
2003b) and Tycho's SNR (V\"olk et al. 2002) such a strong magnetic field
amplification can only be produced as a nonlinear effect by a very
efficiently accelerated {\it nuclear CR component}. Its energy density,
consistent with all existing data, is so high that it is able to strongly
excite magnetohydrodynamic fluctuations, and thus to amplify the upstream
magnetic field $B_\mathrm{ISM}$ to an effective field
$B_\mathrm{0}>B_\mathrm{ISM}$, and at the same time to permit efficient CR
scattering on all scales, approaching the Bohm limit. The same large
effective magnetic field is required by the comparison of our
self-consistent theory with the {\it morphology} of the observed X-ray
synchrotron emission, in particular, its spatial fine structure (Berezhko
et al. 2003a; Berezhko \& V\"olk 2004).

As the simplest possible conclusion from these independent but convergent
considerations we shall assume that a constant fraction $\delta$ of the CR
pressure $P_\mathrm{c}$ at the shock is converted into magnetic field
energy so that $B^2/(8\pi)=\delta \cdot P_\mathrm{c}$. Formally then, the
magnetic field in the upstream preshock medium becomes time dependent,
independent of the magnetic field in the ambient interstellar medium
around the SNR:
\begin{equation}
B_\mathrm{0}=\sqrt{8\pi \delta \cdot P_\mathrm{c}}.
\label{B0}
\end{equation}
We use the moderate parameter value $\delta= 10^{-3}$. In the active SNR
phase, when $P_\mathrm{c}\sim \rho_\mathrm{0} V_\mathrm{s}^2$, this results in an effective
downstream magnetic field of energy density 
\begin{equation}
B_\mathrm{d}^2/(8\pi)\approx 10^{-2}\rho_\mathrm{0} V_\mathrm{s}^2. 
\label {BD}
\end{equation}

Such a large downstream magnetic field value is consistent with the values
derived from the nonthermal X-rays measured in individual SNRs (Berezhko
et al. 2003a; Berezhko \& V\"olk 2004). When in the course of SNR
evolution this field value drops below the magnetic field value
$B_\mathrm{ISM}$ in the ambient ISM, we shall use
$B_\mathrm{0}=B_\mathrm{ISM}=5$~$\mu$G.  Note that this amplified magnetic
field $B_\mathrm{0}$ exceeds the typical ISM value $B_\mathrm{ISM}\approx
5$~$\mu$G during almost the entire evolution except in the very late SNR
evolutionary phase.  Therefore the results of our calculations below are
insensitive to the concrete value of $B_\mathrm{ISM}$.

To a good approximation the amplified field $B_\mathrm{d}(r,t)$ in the
downstream region is spatially uniform and equal to its value
$B_\mathrm{2}(t)$ just behind the shock (see Appendix). This is also
roughly true for the free expansion phase. For simplicity in what follows
we shall make use of this property for all evolutionary epochs considered.

The number of suprathermal protons injected into the acceleration process
is described by a dimensionless injection parameter $\eta$ which is a
fixed fraction of the number of ISM particles entering the shock front.  
For simplicity assume that the injected particles have a velocity equal to
four times the postshock sound speed. Note that the value of the injection
velocity $v_\mathrm{inj}$ is not a significant parameter. Since it divides
the whole particle distribution into two physically different components,
thermal particles and energetic particles (which we call here cosmic
rays), the only significant requirement for the value $v_\mathrm{inj}$ is
that it should be high enough, so that the diffusive approach is valid for
all energetic particles (for a detailed discussion, see Malkov \& V\"olk
1995). For a given value $v_\mathrm{inj}$ the only relevant parameter
which determines the number of suprathermal particles injected into the
acceleration is the injection parameter $\eta$.

Assuming a purely parallel shock, $\eta$ is estimated to be $\eta \approx
10^{-2}$ (e.g. Scholer et al. 1992; Malkov 1998). As analyzed in detail by
V\"olk et al. (2003) the directional structure of the fluctuating magnetic
field at the shock front has the effect of increasingly suppressing the
leakage of suprathermal particles from the downstream region back upstream
when the shock becomes instantaneously more and more oblique.  Applied to
the spherical SNR shock in a large-scale external field $B_\mathrm{ISM}$, the
local injection rate averaged over the fluctuating magnetic field
directions is lower than for a purely parallel shock by a factor $\approx
10^{-2}$, and even this reduced injection takes place only on some
fraction $f_\mathrm{re}=0.15$ to 0.25 of the shock surface, depending on the size
of the SNR. Therefore we adopt here a value $\eta\sim 10^{-4}$ for the
injection parameter and choose a renormalization factor $f_\mathrm{re}=0.2$. The
latter is used to renormalize (reduce) the number of CRs calculated within
our spherically symmetric model. These parameter values, used for the
entire sample of SNRs discussed in this paper, are close to the values
determined individually for the three objects SN~1006, Tycho's SNR, and
Cas~A, mentioned before. In detail the objects in the sample will
naturally be somewhat different. Thus our theoretical analysis is expected
to be appropriate for an entire population, even though it will not give a
precise result for each individual object separately.

Since the proton injection rate is not known more precisely than by a
factor of order unity, we compare the results of our calculations
performed for $\eta=10^{-4}$ with those corresponding to $\eta=3\times
10^{-4}$ in order to study the sensitivity of the calculations to the
value of $\eta$.

It is assumed that also electrons are injected into the acceleration
process at the shock front. Formally their injection momentum is taken to
be the same as that of the protons. Since the details of the electron
injection process are poorly known, we chose the electron injection rate
such that the electron:proton ratio $K_\mathrm{ep}$ (which we define as
the ratio of their distribution functions at all rigidities where the
protons are already relativistic and the electrons have not yet been
cooled radiatively) is a constant to be determined from the observations.

The electron dynamics is exactly the same as that for protons for electron
rigidities corresponding to ultrarelativistic protons, as long as
synchrotron losses are neglected. Therefore, beyond such rigidities and
below the loss region the distribution function of accelerated electrons
has the form
\begin{equation} f_\mathrm{e}(p)= K_\mathrm{ep} f(p)
\label{fe} 
\end{equation} 
at any given time.

In the sequel we shall use a value $K_\mathrm{ep}=10^{-2}$ of the
electron-to-proton ratio, similar to what is observed for the Galactic
CRs. Only for sufficiently large momenta will the electron distribution
function $f_\mathrm{e}(p)$ deviate from this relation as a result of
synchrotron losses. These are taken into account by supplementing the
ordinary diffusive transport equation by a corresponding loss term:
\begin{equation}
{\partial f_\mathrm{e}\over \partial t}=\nabla \kappa \nabla f_\mathrm{e} 
-\vec{w}\nabla f_\mathrm{e}
+\frac{\nabla \vec{w}}{3}p\frac{d f_\mathrm{e}}{dp}
+\frac{1}{p^2}\frac{d}{dp}
\left( \frac{p^3}{\tau_\mathrm{l} }f_\mathrm{e}\right),
\label{eqfe}
\end{equation}
where the first three terms on the right hand side 
describe diffusion, convection due to the mean gas speed $\vec{w}$,
and adiabatic expansion/compression, respectively. The synchrotron loss 
time in the
fourth term is given by the expression (e.g. Berezinskii et al., 
1990)
\begin{equation}
\tau_\mathrm{l}=\left( \frac{4 r_\mathrm{0}^2 B^2p}{9 m_\mathrm{e}^2c^2}\right)^{-1},
\label{taul}
\end{equation}
where $m_\mathrm{e}$ is the electron mass and 
$r_\mathrm{0}$ is the classical electron radius.

The solution of the transport equations for the energetic protons and
electrons, and of the gas dynamic equations at each instant of time
yields the CR spectra and the spatial distributions of CRs and thermal
plasma. This makes it possible to calculate the expected nonthermal emission
produced by CRs in SNRs.

The choice of $K_\mathrm{ep}$ allows one to determine the electron 
distribution function and
to calculate the associated emission. The expected synchrotron 
SNR luminosity is given by the expression (e.g. Berezinskii et al.
1990)
\begin{equation}
L_\mathrm{\nu}=3.8\times 10^{- 20} f_\mathrm{re}
\int_0^{\infty} dr r^2 B_\mathrm{\perp}\int_0^{\infty} dp p^2 f_\mathrm{e}(r,p,t) 
F\left(\frac{\nu}{\nu_\mathrm{c}} \right)
\label{Lnu}
\end{equation}
in erg/(s~Hz), where
\begin{equation}
F(x)=x\int_x^{\infty}K_{5/3}(x')dx',
\label{F}
\end{equation}
$K_{\mu}(x)$ is the modified
Bessel function, $\nu_\mathrm{c}=3e B_\mathrm{\perp} p^2 /[ 4\pi (m_\mathrm{e}c)^3]$, and
$B_\mathrm{\perp}$ is the  magnetic field component
perpendicular to the line of sight. It
yields the flux density $S_\mathrm{\nu}$ at distance $d$:
\begin{equation}
S_\mathrm{\nu}=L_\mathrm{\nu}/(4\pi d^2).
\label{Snu}
\end{equation}

Since in the shock region with efficient particle injection and
acceleration the upstream magnetic field directions can be assumed to be
almost completely randomized due to intense generation of magnetic
fluctuations, we adopt the postshock magnetic field strength
$B_\mathrm{2}=\sigma B_\mathrm{0}$, taking into account that at the shock
front the perpendicular field component undergoes MHD compression as a
result of the gas compression. Below we shall use $B_\mathrm{\perp}=0.5B$.

\section{Supernova dynamics}

At the start our study, we assume for the SN explosion energy a typical
value $E_\mathrm{sn}=10^{51}$~erg. We also restrict our consideration to
the case of a uniform ISM and type Ia SNe which means
$M_\mathrm{ej}=1.4M_{\odot}$ and $k=7$ (see Eq. (1)). Since the
correlation between the SN sites and the ambient ISM density structure is
not known we consider three different phases of the ISM with hydrogen
number densities $N_\mathrm{H}=3$, 0.3 and 0.003~cm$^{-3}$, which
determine the ISM mass density as
$\rho_\mathrm{0}=1.4m_\mathrm{p}N_\mathrm{H}$, where $m_\mathrm{p}$ is the
proton mass. The numerical results for the time evolution of individual
SNRs are shown in Fig.1.

The gas dynamic part of the problem is characterized by the following
length, time and velocity scales:
\[R_\mathrm{0}=(3 M_\mathrm{ej}/ 4\pi \rho_\mathrm{0})^{1/3},~ t_\mathrm{0}=R_\mathrm{0}/V_\mathrm{0},~
V_\mathrm{0}=\sqrt{2E_\mathrm{sn}/{M_\mathrm{ej}}},
\]
which correspond to the sweep-up radius, the sweep-up time and the mean
initial speed of the ejecta, respectively.  The scaling values for the cases
considered are $R_\mathrm{0}=1.45$, 3.2, 14.9~pc and $t_\mathrm{0}=166$, 367, 1709~yr,
respectively.

We note that the following simplified approach is frequently used for the
description of SNR dynamics. It treats the ejecta as initially
expanding as a whole with the single speed $V_\mathrm{0}$. In this case two
different SNR evolutionary phases are distinguished: a free expansion
phase with roughly constant shock speed $V_\mathrm{s}\approx V_\mathrm{0}$ (which lasts up
to $t\approx t_\mathrm{0}$) and the subsequent selfsimilar Sedov phase for
$t>t_\mathrm{0}$.

The actual ejecta, created after the SN explosion, have a broad
distribution in velocity $v$, cf. Eq. (1). Due to this fact the SN shock
undergoes substantial deceleration not only at $t>t_\mathrm{0}$ but also at
$t<t_\mathrm{0}$.  Despite this logical contradiction we shall still use below
the term ``free expansion phase'' for the SNR evolution period
corresponding to $t<t_\mathrm{0}$.

As can be seen from Fig.1, the shock speed is nevertheless constant during
a short initial period. This occurs because we restrict the velocity
distribution of the ejecta to a maximum speed $v_\mathrm{max}=4\times
10^4$~km/s, which is also the initial speed of the outer part of the
ejecta (piston) $V_\mathrm{i}$. This period lasts up to the instant of
time when the swept-up mass
\begin{equation}
M_\mathrm{sw}=(4\pi/3)R_\mathrm{s}^3\rho_\mathrm{0}
\label{Msw}
\end{equation}
becomes equal to that part of the mass of the ejecta which has the initial
speed $v=v_\mathrm{max}$. During this initial period the shock speed is
$V_\mathrm{s}=1.1v_\mathrm{max}$.

During the subsequent part of the free expansion phase ($t<t_\mathrm{0}$) the
shock expansion law is
\begin{equation}
R_\mathrm{s} \propto
  E_\mathrm{sn}^\mathrm{(k-3)/2k}M_\mathrm{ej}^\mathrm{(5-k)/2k} 
\rho_\mathrm{0}^\mathrm{-1/k} t^\mathrm{(k-3)/k}
\label{Rsfe}
\end{equation}
(Chevalier, 1982) which for $k=7$ gives
\begin{equation}
R_\mathrm{s}\propto [E_\mathrm{sn}^2/(M_\mathrm{ej} \rho_\mathrm{0})]^{1/7}t^{4/7}.
\label{Rsk7}
\end{equation}

In the following adiabatic (Sedov) phase $(t\gsim t_\mathrm{0})$ we have 
\begin{equation}
R_\mathrm{s}\propto (E_\mathrm{sn}/
\rho_\mathrm{0})^{1/5}t^{2/5}.
\label{Rssed}
\end{equation}
%
%
%------------------------------------------------------------------------fig.1-
\begin{figure}
\centering
\includegraphics[width=7.5cm]{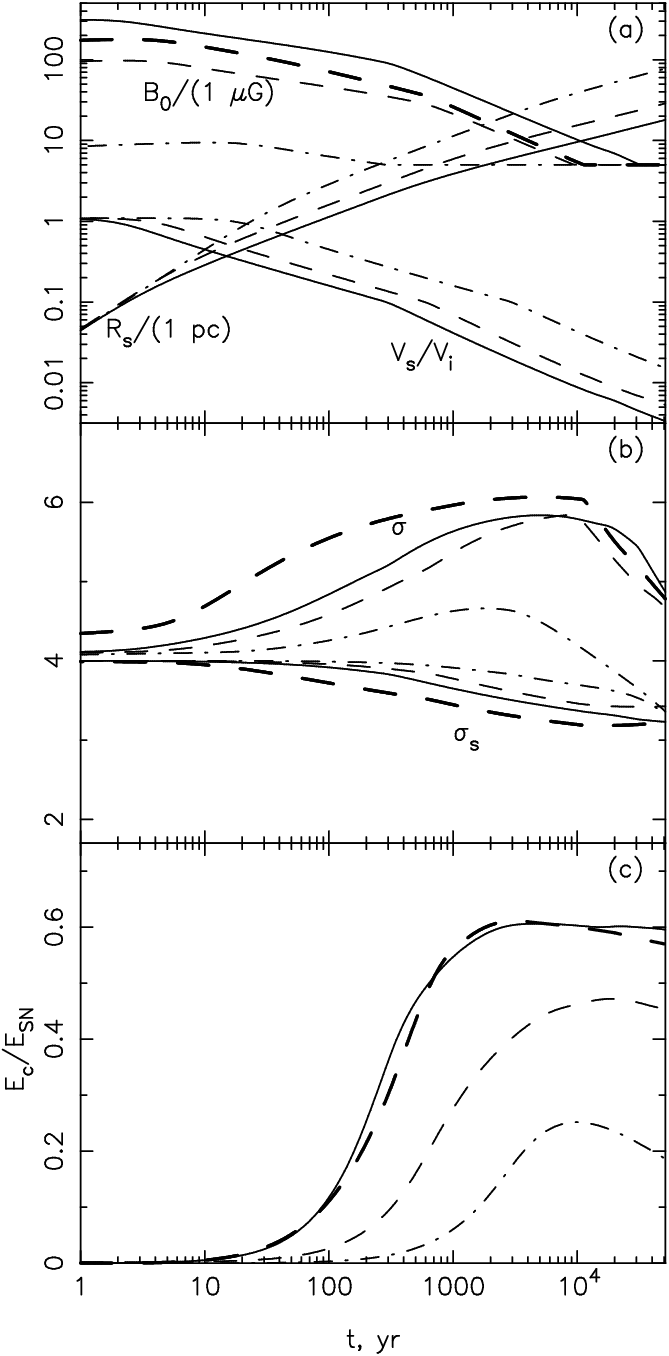}
\caption{Upstream effective magnetic field $B_\mathrm{0}$, 
shock radius $R_\mathrm{s}$ and shock speed $V_\mathrm{s}$ (a); total shock ($\sigma$)
and subshock ($\sigma_\mathrm{s}$) compression ratios (b); CR
energy $E_\mathrm{c}$ as a function of time for the case of SN explosion energy
$E_\mathrm{sn}=10^{51}$~erg and ISM gas number densities
$N_\mathrm{H}=3$~cm$^{-3}$ (solid lines), 0.3~cm$^{-3}$ (dashed lines) and
0.003~cm$^{-3}$ (dash-dotted lines) in the spherically symmetric 
case (c). $V_\mathrm{i}=4\times 10^4$~km/s is the initial piston speed.
Thick dashed lines correspond to the case $E_\mathrm{sn}=10^{51}$~erg, 
$N_\mathrm{H}=0.3$~cm$^{-3}$
with higher injection rate $\eta=3\times 10^{-4}$.}
\end{figure}
%------------------------------------------------------------------------------
%

The decelerating shock becomes progressively more modified, cf. Fig.1b.
The degree of shock modification is very well illustrated by the behavior
of the shock compression ratios. When the CR pressure $P_\mathrm{c}$ is
small compared to the ram pressure $\rho_\mathrm{0}V_\mathrm{s}^2$ the
shock is unmodified and has a compression ratio close to the classical
value $\sigma=4$.  When $P_\mathrm{c}$ becomes comparable to
$\rho_\mathrm{0}V_\mathrm{s}^2$, CR backreaction leads to an increase of
the total shock compression ratio $\sigma$ , whereas the subshock
compression ratio $\sigma_\mathrm{s}$ decreases. During an initial
evolution period $t<10^3$~yr the total shock compression ratio goes to the
value, expected for a strongly modified shock, (Berezhko \& Ellison 1999)
\begin{equation}
\sigma \approx 1.5 M_\mathrm{a}^{3/8},
\label{sigma}
\end{equation}
where $M_\mathrm{a}=V_\mathrm{s}/c_\mathrm{{a0}}$ is the shock Alfv\'(e)n
Mach number, $c_\mathrm{{a0}}=B_\mathrm{0}/\sqrt{4\pi \rho_\mathrm{0}}$ is
the Alfv\'(e)n speed. This limit is achieved at the stage $t\approx
3\times 10^3$~yr. For example, in the case of $N_\mathrm{H}=3$~cm$^{-3}$,
$M_\mathrm{a}\approx 30$ and $\sigma_\mathrm{max}\approx 6$, in good
agreement with the numerical results (Fig.1b).

Since for $t=10^3 - 10^4$~yr the amplified field $B_\mathrm{0}\propto
V_\mathrm{s}$, the Alfv\'(e)n Mach number remains roughly constant.
Therefore the shock modification is also nearly constant during this
period of time (see Fig.1b). When $B_\mathrm{0}$ eventually becomes equal
to $B_\mathrm{ISM}$ the Alfv\'en Mach number $M_\mathrm{a}\propto
V_\mathrm{s}$ decreases and, according to Eq.(\ref{sigma}), the shock
compression ratio $\sigma\propto V_\mathrm{s}^{3/8}$ also diminishes in
time.

The non-renormalized CR energy content
\begin{equation}
E_\mathrm{c}(t)=16\pi^2 \int_0^{\infty}dr r^2 \int_0^{\infty}dp p^2
\epsilon_\mathrm{k} f(r,p,t),
\label{Ec}
\end{equation}
where 
$\epsilon_\mathrm{k}=\sqrt{p^2c^2+m_\mathrm{p}^2c^4}-m_\mathrm{p}c^2$ is
the proton kinetic energy, increases during the SNR evolution up to the
time $t\sim 10^4$~yr, when the shock compression ratio starts to decrease
rapidly to values $\sigma \sim 4$ and even less (Fig 1c).

When the SNR age approaches $t\approx 10^5$~yr, the CR acceleration
efficiency also decreases as a result of the fall-off of the Alfv\'en Mach
number $M_\mathrm{a}$. Due to its relatively low speed the shock at this stage
continues to produce CRs albeit with relatively low cutoff momentum
$p_\mathrm{m}(t)$, whereas previously produced CRs with $p>p_\mathrm{m}$ leave the shock
volume since their diffusive propagation is now
faster than that of the previously confining shock. This process of CR
escape from the shock volume can be even faster in the case of strong
Alfv\'en wave damping as a result of their nonlinear interaction (Ptuskin \&
Zirakashvili 2003). Therefore this late stage of the SNR evolution,
$t\gsim 3\times 10^4$~yr, does not play an important role in CR
production, even though such an old SNR may still confine low energy CRs
and shine in the radio range.

\section{The accelerated CR distribution}
\subsection{Proton distribution}
To illustrate the evolution of the accelerated CR spectrum we consider 
the case of $N_\mathrm{H}=0.3$~cm$^{-3}$ and $E_\mathrm{sn}=10^{51}$~erg.
%
%------------------------------------------------------------------------fig.2-
\begin{figure}
\centering
\includegraphics[width=7.5cm]{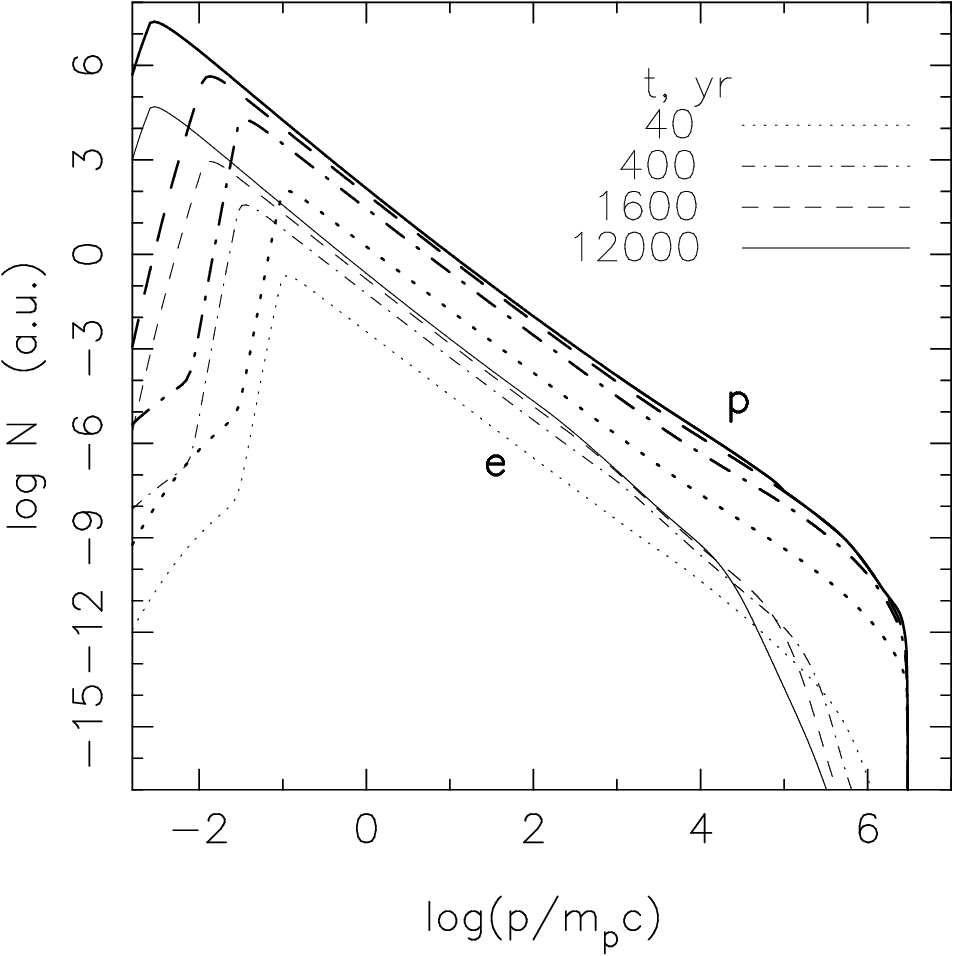}
\caption{The overall CR momentum spectrum for 
four
different times, for the case of
$E_\mathrm{sn}=10^{51}$~erg and $N_\mathrm{H}=0.3$~cm$^{-3}$. Thick and thin lines
correspond to protons and electrons respectively.}
\end{figure}
%------------------------------------------------------------------------------
The volume-integrated (or overall) CR distribution
\begin{equation} 
N(p,t)=16\pi^2p^2 \int_0^{\infty}dr r^2 f(r,p,t) 
\label{N} 
\end{equation}
for the CR proton and CR electron components is presented in Fig.2 for 
different times. The amplitude of the spectrum
increases with time roughly proportional to the CR energy content
$E_\mathrm{c}$. This process becomes very slow  
for $t>10^3$~yr as seen from Fig.2.

The maximum (or cutoff) momentum of the proton spectrum at a given
evolutionary stage is determined by geometrical factors and can be 
estimated by the expression (Berezhko 1996)
\begin{equation}
\frac{p_\mathrm{m}}{m_\mathrm{p}c}=\frac{R_\mathrm{s}V_\mathrm{s}}{A\kappa(m_\mathrm{p}c)},
\label{pm}
\end{equation}
where
\begin{equation}
A=4[2+2b+e-(\nu-1)/\nu+d].
\label{A}
\end{equation}
The dimensionless parameters $\nu, e, b$ and $d$ determine the time
variation of the shock size $\nu=d\ln R_\mathrm{s}/d\ln t$, of the diffusion
coefficient $e=d\ln \kappa/d\ln t$, of the injection momentum $b=d\ln
p_\mathrm{inj}/d\ln R_\mathrm{s}$, and of a measure of the inverse compression ratio
$d=[2V_\mathrm{s}-u+R_\mathrm{s}(du/dr)]_2/(u_1-u_2)$.  Here $u$ is the plasma speed with
respect to the shock front and the subscripts 1(2) correspond to the
points just ahead (behind) the subshock front.

Let us consider the two SNR evolutionary periods when the SN
shock is already strongly modified, i.e. the end of the free expansion
phase and the Sedov phase. For a rough estimate one can take $P_\mathrm{c}=0.5
\rho_\mathrm{0}V_\mathrm{s}^2$. In the free expansion phase $\nu=4/7$, $e=3/7$, $b=-3/4$,
$d=-1/2$.  This gives $A=5$ and
\begin{equation}
\frac{p_\mathrm{m}}{m_\mathrm{p}c}=3.6\times 10^6
\left( \frac{V_\mathrm{s}}{10^4~\mbox{km/s}}\right)^{2/3}.
\label{pmfe}
\end{equation}
In the Sedov phase we have $\nu=2/5$, $e=3/5$, $b=-3/2$, $d=3$, which give
$A=16$ and
\begin{equation}
\frac{p_\mathrm{m}}{m_\mathrm{p}c}=1.1\times 10^6
\left( \frac{V_\mathrm{s}}{10^4~\mbox{km/s}}\right)^{4/3}.
\label{pmsed}
\end{equation}
The results presented in Fig.1a and Fig.2 agree very well with these
estimates.

Since the most energetic CRs in the power law part of their spectrum
are produced at the very end of the free expansion phase, that is at
$t\sim t_0$, the maximum CR momentum depends on physical parameters as 
$p_\mathrm{max}\propto V_0 R_0 B_0(t_0)$. This gives
\begin{equation}
\frac{p_\mathrm{max}}{m_\mathrm{p}c}\approx
10^6 \left( \frac{E_\mathrm{sn}}{10^{51}~\mbox{erg}}\right)
\left(\frac{M_\mathrm{ej}}{1.4M_{\odot}}\right)^{-2/3}
\left(\frac{N_\mathrm{H}}{0.3~\mbox{cm}^{-3}}\right)^{1/6},
\label{pmax}
\end{equation}
where we have explicitly given the dependence on the SN parameters.  The
maximum CR momentum depends strongly on $E_\mathrm{sn}$ and
$M_\mathrm{ej}$, but is only weakly dependent upon the ISM density
$\rho_0$.

We note that a maximum proton momentum $p_\mathrm{max}\sim
10^6m_\mathrm{p}c$ is required to reproduce the spectrum of Galactic CRs
up to the knee energy (Berezhko \& Ksenofontov 1999).

The maximum momentum in the overall CR spectrum, which is almost constant
at the late epoch $t> 10^3$~yr, is about $p_\mathrm{max}\approx
10^6~m_\mathrm{p}c$ and corresponds to the beginning of the Sedov phase.
All the very high-energy protons produced during the preceding free
expansion phase $t<t_0$ contribute only a very steep power law tail at
larger momenta $p>p_\mathrm{max}$ in the final overall spectrum. The shape
of this high energy tail can be derived as follows:

The kinetic energy of the ejecta contained in the fraction with speed
$v$ within the interval $dv$ is
\begin{equation}
dE_\mathrm{ej}(v)\propto v^\mathrm{4-k}dv.
\label{Eej}
\end{equation}

The CR spectrum created by the strongly modified shock is quite hard:
$N(p)\propto p^{-\gamma}$, $\gamma <2$. For the sake of a rough estimate
we assume that all the above energy of the ejecta goes into CRs with
maximum momenta $p_\mathrm{m}(V_\mathrm{s})$, where the shock speed is
roughly the speed $V_\mathrm{s}\approx v$ of the ejecta. 
This gives
\begin{equation}
N(p_\mathrm{m})p_\mathrm{m}dp_\mathrm{m}\propto dE_\mathrm{ej}
\label{NEej}
\end{equation}
in the ultrarelativistic case.
During the period of SNR evolution when the shock is already modified
we have $B_0\propto V_\mathrm{s}$ and therefore $p_\mathrm{m}\propto R_\mathrm{s}V_\mathrm{s}B\propto
V_\mathrm{s}^{2-\nu/(1-\nu)}$ so that we obtain $N\propto p^{-\gamma}$ with
\begin{equation}
\gamma=(9k-33)/6.
\label{gamma}
\end{equation}

For the case $k=7$ this gives $\gamma=5$ which agrees with the numerical
results (Fig.2). This consideration shows the following: despite the fact
that during the free expansion phase the SN shock produces CRs with a
power law spectrum up to the momentum $p_\mathrm{m}(t_0)$ which is larger
than what is attained in the Sedov phase, $p_\mathrm{m}(t>t_\mathrm{0})$,
these particles are seen only in a very steep part of the final overall CR
spectrum due to their low number (low energy content).  In addition,
significant shock modification does not start at the very beginning of SNR
evolution but only at $t\gsim 100$~yr. Due to the fact that efficient CR
production (followed by strong shock modification and magnetic field
amplification) takes place only during a relatively short period in
the free expansion phase, this power law spectrum ends in an exponential
cutoff at a momentum which is only a factor of 3 larger than
$p_\mathrm{max}$. Basically these particles play no role for the
population of Galactic CRs. We should add here that in the case of type II
and Ib SNe, which are more numerous in our Galaxy (Tammann et al. 1994),
the value of the parameter $k$ is even larger, $k\approx 10$ (e.g.
Chevalier \& Fransson 1984). In these cases the free expansion phase
presumably contributes even less to the final CR spectrum.

\subsection{Electron distribution} 
At low momenta we have
$N_\mathrm{e}(p)\propto N(p)$. However, the shape of the overall electron
spectrum deviates from that of the proton spectrum at high momenta
$p>p_\mathrm{l}\approx 10^3m_\mathrm{p}c$ due to the synchrotron losses in
the downstream magnetic field which is assumed uniform
($B_\mathrm{d}=B_2=\sigma B_0$). According to expression (6) the
synchrotron losses become important for electron momenta greater than
\begin{equation} 
\frac{p_\mathrm{l}}{m_\mathrm{p}c} \approx \left(\frac{10^8~\mbox{yr}}{t}\right)
\left(\frac{10~\mu\mbox{G}}{B_\mathrm{d}}\right)^2. 
\label{pl}
\end{equation}

The shock continues to produce an electron spectrum $f_\mathrm{e}\propto p^\mathrm{-q}$ 
with $q\approx 4$ up
to a maximum momentum $p_\mathrm{max}^\mathrm{e}$ which is significantly larger than
$p_\mathrm{l}$. In the momentum range from $p_\mathrm{l}$ to
$p_\mathrm{max}^\mathrm{e}$, due to the synchrotron losses the overall electron spectrum
is therefore close to $N_\mathrm{e}\propto p^{-3}$.

The maximum electron momentum can be estimated by equating the 
synchrotron loss time with the acceleration time. This results in
(e.g. Berezhko et al. 2002)
\[
\frac{p_\mathrm{max}^e}{m_\mathrm{p}c}
= 6.7\times 10^4 \left(\frac{V_\mathrm{s}}{10^3~\mbox{km/s}}\right)
\]
\begin{equation}
\hspace{1cm}\times
\sqrt{\frac{(\sigma-1)}{\sigma (1+ \sigma^2)}
  \left(\frac{10~\mu\mbox{G}}{B_0}\right)}.
  \label{pemax} 
\end{equation}

The main fraction of the electrons with the highest energies $p\sim
10^5m_\mathrm{p}c$ is produced at the end of the free expansion phase. In
this stage, according to Fig.1a, $V_\mathrm{s}\sim V_0=8.5\times
10^3$~km/s and $B_0\approx 30$~$\mu$G, which leads to a maximum electron
momentum $p_\mathrm{max}^\mathrm{e}\approx 4\times 10^5m_\mathrm{p}c$, in
agreement with the numerical results (Fig.2).

\section{Synchrotron luminosity}
To illustrate the time  variation of the synchrotron spectrum during the 
SNR
evolution we present in Fig.3 the calculated synchrotron luminosity 
$L_\mathrm{\nu}(\nu)$
in an ISM with number density $N_\mathrm{H}=0.3$~cm$^{-3}$,
for different ages.
The shape of the spectrum $L_\mathrm{\nu}(\nu)$ is directly related to the
overall electron spectrum $N_\mathrm{e}(p)$, shown in Fig.2.
For relatively low frequencies $\nu<10^{14}$~Hz it has a power law form 
\begin{equation}
L_\mathrm{\nu}\approx A_\mathrm{\nu}B^\mathrm{\alpha+1}\nu^{-\alpha},
\label{L(Bnu)}
\end{equation}
where the power law index
$\alpha=(\gamma-1)/2$ is related to the power law index $\gamma$ which
determines the electron spectrum $N_\mathrm{e}\approx A_\mathrm{e}p^{-\gamma}$ near the
electron momentum 
\begin{equation}
\frac{p}{m_\mathrm{e}c}= \left(\frac{4\pi m_\mathrm{e}c^2 \nu}{0.87 
ecB_{\perp}}\right)^{1/2},
\label{pe(nu)}
\end{equation}
which gives the maximum contribution to the synchrotron emission at
frequency $\nu$.
Due to the shock modification the electron spectrum has a concave shape
characterized by an index $\gamma$ which slowly decreases with increasing 
energy. Therefore the
synchrotron spectrum also has a concave shape with $\alpha>0.5$ at the
lowest frequencies $\nu\sim 1$~GHz, decreasing to $\alpha=0.5$ at $\nu
\gg 1$~GHz (e.g. Reynolds \& Ellison 1992; Berezhko et al. 2002).

The current luminosity $L_\mathrm{\nu}$ is determined by the downstream
magnetic field value at the current epoch $B_\mathrm{d}\approx \sigma
B_0$, which decreases with time due to the shock deceleration, and by the
total number of electrons produced during all the previous stages
$A_\mathrm{\nu}\propto A_\mathrm{e}$. During the initial period,
$t<10^3$~yr, the factor $A_\mathrm{\nu}\propto A_\mathrm{e}$ grows more
rapidly than the factor $B^\mathrm{\alpha+1}$ decreases and this leads to
an increase of $L_\mathrm{\nu}$ in this period. In the later epoch
$t>2\times 10^3$~yr the number of accelerated electrons increases only
slowly with time. Therefore the decrease of the magnetic field strength
leads to a decrease of $L_\mathrm{\nu}$.

The synchrotron emission at the highest frequencies $\nu > 10^{14}$~Hz is
produced by the electrons with $p>p_\mathrm{l}$, which suffer dramatic
synchrotron cooling. Therefore, in addition to the above physical factors,
the behavior of the synchrotron spectrum $L_\mathrm{\nu}$ in this
frequency range is influenced by the values of $p_\mathrm{l}(t)$ and
$p_\mathrm{max}^\mathrm{e}(t)$. Taking into account expression
(\ref{pemax}) for the electron cutoff momentum and relation
(\ref{pe(nu)}), one can easily derive the value of the cutoff frequency
\begin{equation}
\nu_\mathrm{max}=2.1\times 10^{17} b_{\perp} \frac{\sigma-1}{\sigma^2+1}
\left(\frac{V_\mathrm{s}}{10^3~\mbox{km/s}}\right)^2~\mbox{Hz},
\label{numax}
\end{equation}
where $b_{\perp}=B_\mathrm{{d}\perp}/B_\mathrm{d}$. It is taken as $b_{\perp}=0.5$ in our 
calculation. Since the cutoff frequency is mainly determined by the
shock speed, its value $\nu_\mathrm{max}$ goes down quickly during the 
SNR
evolution due to the shock deceleration, as seen from 
Fig.3. 
%
%------------------------------------------------------------------------fig.3-
\begin{figure}
\centering
\includegraphics[width=7.5cm]{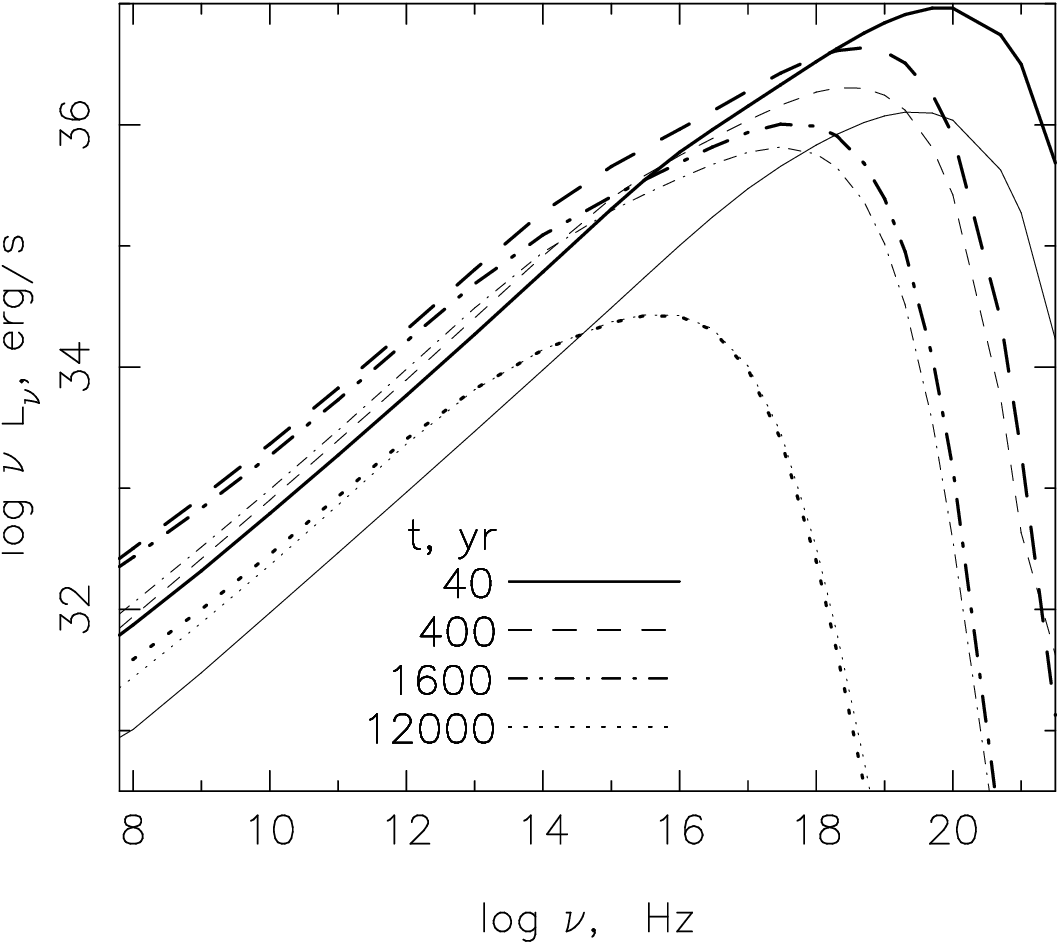}
\caption{Synchrotron spectral luminosity distribution as a function of
frequency at four different times, for the same case as in Fig.2 (thin
lines). Thick lines correspond to the injection parameter value
$\eta=3\times 10^{-4}$.}
\end{figure}
%------------------------------------------------------------------------------
It is interesting to note that at frequency $\nu\sim 10^{18}$~Hz,
corresponding to the typical hard X-ray energy
$\epsilon_\mathrm{\nu}=4$~keV, only relatively young SNRs of age $t \lsim
10^3$~yr produce very intense nonthermal X-ray emission which belongs to
the power law part of the synchrotron spectrum $L_\mathrm{\nu}(\nu)$. Over
the subsequent epochs $t \gsim 10^3$~yr the shock speed becomes so small
that the above X-ray frequency range falls into to the exponential tail of
the spectrum. This leads to a fast decrease of the nonthermal X-ray
emission during this period of SNR evolution.

To illustrate the evolution of the synchrotron flux we
present in Fig.4 the luminosity $L_\mathrm{\nu}$ at $\nu=1$~GHz 
as a function of SNR diameter for the case of SNRs with explosion energy
$E_\mathrm{sn}/(10^{51}~\mbox{erg})$ = 0.25, 1 and 3 in an ISM of density 
$N_\mathrm{H}/(1~\mbox{cm}^{-3})$ = 0.003, 0.3 and 3.

The variation of the synchrotron flux is due to variation of
$R_\mathrm{s}$, $V_\mathrm{s}$ and $B$.

During a short initial period, which lasts from $t\approx 1$~yr for
$N_\mathrm{H}=3$~cm$^{-3}$ to $t\approx 10$~yr for
$N_\mathrm{H}=0.003$~cm$^{-3}$ (see Fig.1a), the SN shock speed is constant.
Since the SNR shock 
is not significantly modified in the free expansion phase (see Fig.1b), 
the CR 
distribution function at the shock front has the form
\begin{equation}
f_\mathrm{s}=\frac{4\eta N_\mathrm{H}}{4\pi p_\mathrm{inj}^3}\left(
\frac{p}{p_\mathrm{inj}}
\right)^{-4},
\label{fs}
\end{equation}
which shows that the CR
number density at
the shock front is proportional to
the momentum $p_\mathrm{inj}\propto V_\mathrm{s}$ of the suprathermal particles 
injected at the shock front
into the acceleration process. Therefore we have 
$A_\mathrm{e}\propto N_\mathrm{H} V_\mathrm{s}R_\mathrm{s}^3$, $B_\mathrm{d}\propto \sqrt{N_\mathrm{H}}$ and
\begin{equation}
L_\mathrm{\nu}\propto N_\mathrm{H}^{7/4} D^3,
\label{Lnuin}
\end{equation}
taking into account that in the case of an unmodified shock $\gamma=2$ and
$\alpha=0.5$. This explains the rising part of the calculated curves
$L_\mathrm{\nu}(D)$ which corresponds to small diameter values $D$ in
Fig.4.

During the subsequent part of the free expansion phase the SN shock
decelerates. 
According to Eq.(\ref{Rsk7}) its speed as a function of shock radius is
\begin{equation}
V_\mathrm{s}\propto
[E_\mathrm{sn}^2/(M_\mathrm{ej}N_\mathrm{H})]^{1/4}R_\mathrm{s}^{-3/4}.
\label{Vs(Rs)}
\end{equation}
Since the shock, as in the previous epoch, 
is not yet strongly modified, the CR pressure goes like 
$P_\mathrm{c}\propto N_\mathrm{H} V_\mathrm{s}$, leading to a
magnetic field variation $B_\mathrm{d}\propto \sqrt{N_\mathrm{H} V_\mathrm{s}}$.
The expected luminosity dependence on the SNR diameter is then
\begin{equation}
L_\mathrm{\nu}\propto (E_\mathrm{sn}^2/M_\mathrm{ej})^{7/16}N_\mathrm{H}^{21/16}D^{27/16},
\label{Lnufe}
\end{equation}
which rather well explains the $L_\mathrm{\nu}(D)$ dependence for $D<D_0=2R_0$.
 
In the subsequent Sedov phase a substantial fraction of the total energy
$E_\mathrm{sn}$ goes into the CR component whose overall number remains
nearly constant so that $A_\mathrm{e}\propto E_\mathrm{sn}$.  Note that
the ISM density $N_\mathrm{H}$ influences the amplified magnetic field
value $B_0$, which in turn influences the shock modification and CR
acceleration efficiency (see Fig.1). But this influence is not very
significant: as one can see from Fig.1 an increase of $N_\mathrm{H}$ by a
factor of $10^3$ leads to an increase of $E_\mathrm{c}$ by a factor of
only two. Therefore, for a rough estimate, $E_\mathrm{c}$ and
$A_\mathrm{e}\propto E_\mathrm{c}$ can be considered as independent of
$N_\mathrm{H}$. In this stage the SN shock is significantly modified. This
is a consequence of the fact that the CR pressure is an important fraction
of the shock ram pressure, with $P_\mathrm{c}\propto
\rho_\mathrm{0}V_\mathrm{s}^2$.  The downstream magnetic field
$B_\mathrm{d}\propto \sigma \sqrt{P_\mathrm{c}}$ also varies due to the
variation of the shock compression ratio $\sigma$. Since the variation of
$\sigma$ is much weaker than that of
$P_\mathrm{c}$, we can neglect it for qualitative analysis, taking
$B_\mathrm{d}\propto \sqrt{P_\mathrm{c}}\propto
\sqrt{N_\mathrm{H}}V_\mathrm{s}$. Then, taking into account the expansion
law (\ref{Rssed}) we have
\begin{equation}
L_\mathrm{\nu}\propto E_\mathrm{sn}^{7/4}D^{-9/4}
\label{Lnused}
\end{equation}
independently of $N_\mathrm{H}$ and $M_\mathrm{ej}$, if for simplicity we
again assume the electron spectrum $N_\mathrm{e}\propto p^{-\gamma}$ with
power law index $\gamma=2$, which implies $\alpha=0.5$.  The calculated
electron spectra corresponding to the Sedov phase are steeper due to the
shock modification. This implies $\gamma>2$ at those energies which are
responsible for the radio emission. In addition, the power law index
$\gamma$ varies during shock evolution and therefore the dependencies
$L_\mathrm{\nu}(D)$ are not of a pure power law nature, as is seen from
Fig.4, especially in the case of a dense ISM with $N_\mathrm{H}\gsim
0.1$~cm$^{-3}$ when the shock modification is most significant.  
Nevertheless the above analytical relation can be used for rough
estimates.

In the late Sedov phase, when the effective magnetic field drops to
$B_0=B_\mathrm{ISM}$ and remains constant thereafter, the expected SNR
luminosity $L_\mathrm{\nu}\propto E_\mathrm{sn}$ is roughly independent of
$D$.

%
%------------------------------------------------------------------------fig.4-
\begin{figure}
\centering
\includegraphics[width=7.5cm]{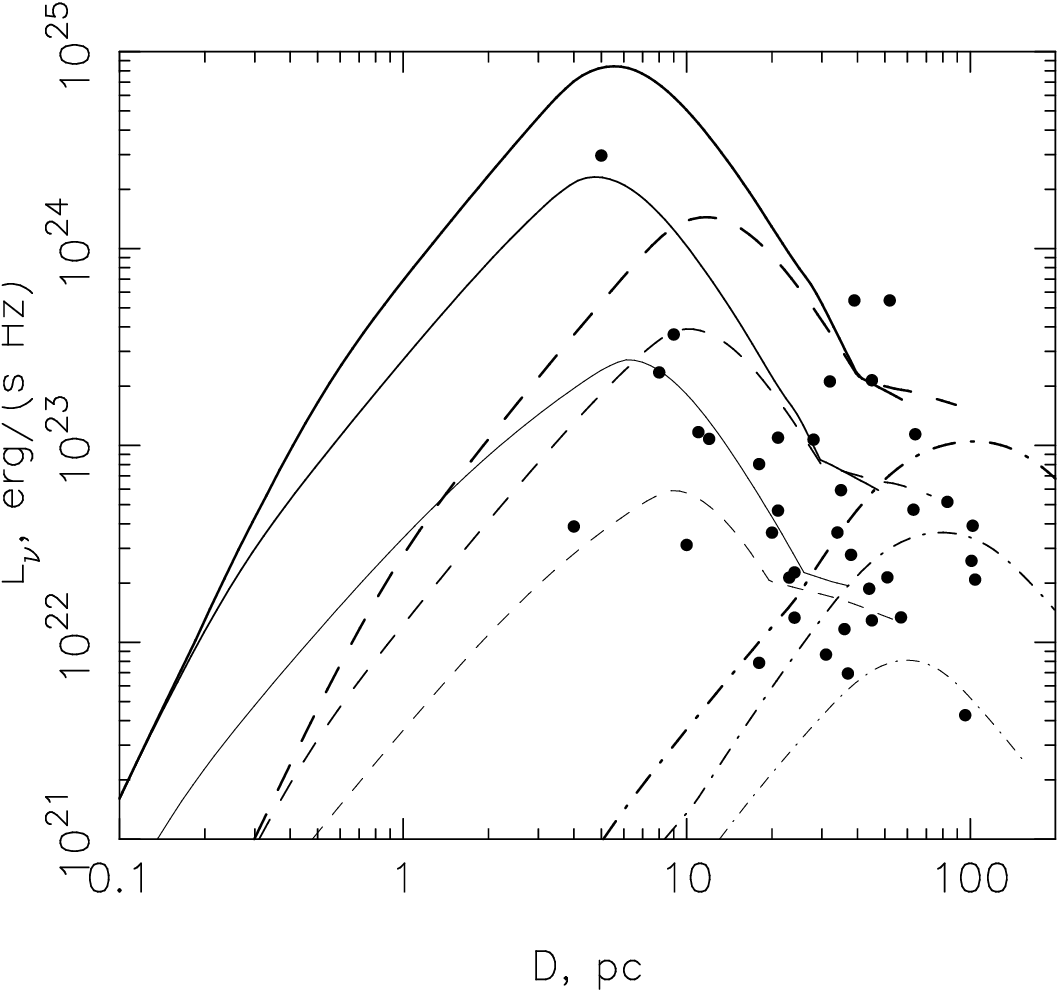}
\caption{The radio spectral luminosity distribution at $\nu=1$~GHz as a 
function of SNR 
diameter in the case of SNR with explosion energy
$E_\mathrm{sn}/(10^{51}~\mbox{erg})$=0.25 (thin lines), 1 (normal) and 3
(thick) 
in an ISM of density $N_\mathrm{H}/(1~\mbox{cm}^{-3})$=3 (solid lines), 0.3
(dashed) and 0.003 (dash-dotted).
Observational data, comprising 37 Galactic SNRs of known distances, are shown
(Case \& Bhattacharya 1998).}
\end{figure}
%------------------------------------------------------------------------------
%
As pointed out by Case \& Bhattacharya (1998), there exists no
significant correlation between radio luminosity and linear diameter.  If
SNe are more or less uniformly distributed over the considered range of
ISM densities then it is clear from Fig.4 that no significant correlation
between $L_\mathrm{\nu}$ and $D$ is to be expected. At least the sample of Case
\& Battacharya appears to be uniformly distributed.

According to the calculated dependence of $L_\mathrm{\nu}$ on $D$ almost all
observed SNRs are at the very end of their free expansion phase or in the
Sedov phase. If this is generally so, then the lack of data
with small SNR sizes $D$ which correspond to the ages $t\ll t_0$
is a consequence of the
relatively low expected luminosity of SNRs of age $t\ll t_0$. In
addition, it can also be due to a selection effect: SNRs with small size
and low luminosity are more difficult to recognize.

At high frequencies $\nu\sim 10^{18}$~Hz, which correspond to the hard
X-ray range, the decrease of the cutoff frequency $\nu_\mathrm{max}$ becomes
appreciable already for $t\gsim 100$~yr. This is why the luminosity
$L_\mathrm{\nu}(D)$ at $\nu \ge 10^{18}$~Hz, shown in Fig.5, 
begins to decrease
already in the early Sedov phase. This factor becomes dominant for
$t>10^3$~yr and leads to a very steep dependence $L_\mathrm{\nu}(D)$. For this
reason only relatively young SNRs in free expansion or in the early Sedov
phase are likely to be detected in nonthermal X-rays. As is clear from
Fig.5, the relatively high synchrotron luminosity in X-rays, expected
already during the free expansion phase, can help to identify
very young SNRs that are not yet bright radio sources.
%
%------------------------------------------------------------------------fig.5-
\begin{figure}
\centering
\includegraphics[width=7.5cm]{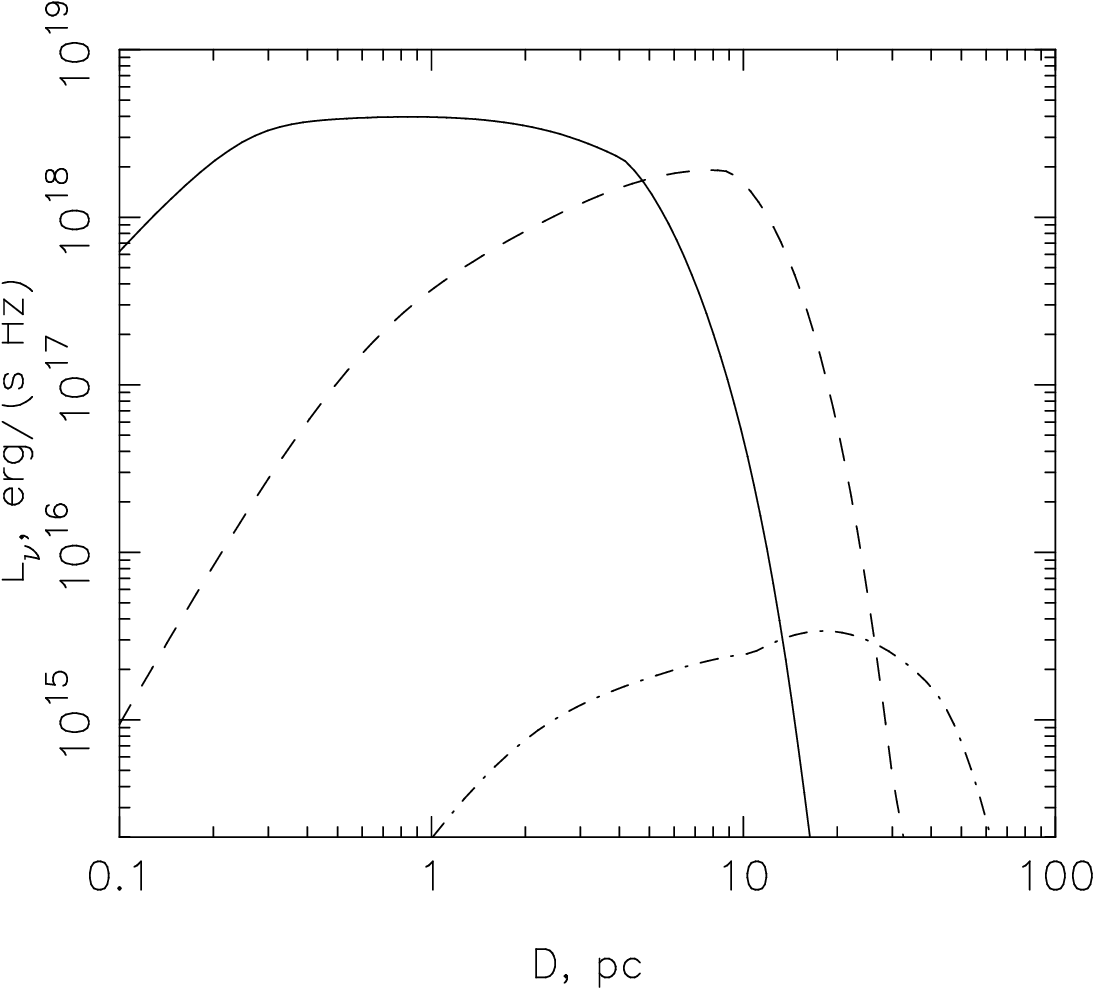}
\caption{The X-ray spectral luminosity distribution at $\nu=10^{18}$~GHz 
(photon energy $\epsilon_\mathrm{\nu}=4$~keV) as a function of SNR 
diameter in the case of a SNR with explosion energy
$E_\mathrm{sn}=10^{51}$~erg
in an ISM of density $N_\mathrm{H}/(1~\mbox{cm}^{-3})$ = 3 (solid line), 0.3
(dashed line) and 0.003 (dash-dotted line).}
\end{figure}
%------------------------------------------------------------------------------
% 

In order to study the sensitivity of our results to the injection rate we
present in Fig.1 and Fig.3 for the case $E_\mathrm{sn}=10^{51}$~erg
$N_\mathrm{H}=0.3$~cm$^{-3}$ the calculations that correspond to the
substantially higher injection rate $\eta =3\times 10^{-4}$. The influence
of $\eta$ is quite different in different SNR evolutionary stages. When
the SN shock is not modified (linear CR acceleration regime), the shape of
the CR spectrum does not depend on $\eta$ and the amplitude of the
spectrum is proportional to $\eta$. Such a situation results during the
initial part of the free expansion phase: as can be seen from Fig.1b, at
$t\lsim 10$~yr the shock is only slightly modified. Since $P_\mathrm{c}\propto
\eta$, the increase of $\eta$ leads to an increase of the magnetic field
value $B_0\propto \sqrt{\eta}$ (Fig.1a), and therefore the synchrotron
luminosity (Fig.3) increases by a factor of about 7 compared with the case
$\eta =10^{-4}$.

When the shock is already strongly modified, i.e. at the end of free
expansion and in the Sedov phase, the relation between $\eta$ and the
characteristics of CRs {\bf in} a SNR become more complicated. The shock
modification and CR energy content, which are closely connected with each
other, are relatively weakly sensitive to the injection rate as can be
seen from Fig.1.  Similarly, the synchrotron luminosities corresponding to
the two values of $\eta$ considered progressively approach each other for
$t\gsim 100$~yr, so that during the Sedov phase $t>10^3$~yr the influence
of $\eta$ on $L_\mathrm{\nu}(\nu)$ becomes very small (see Fig.3). Note
also that for a higher injection rate that leads to larger shock
modification the synchrotron spectrum in the radio range becomes 
steeper due to the concave shape of the spectrum of accelerated CRs.

\section{The $\Sigma-D$ relations in the radio range}
In Fig.6 we present the surface brightness 
\begin{equation}
\Sigma_\mathrm{\nu}=L_\mathrm{\nu}/(\pi^2 D^2)
\label{Sigma}
\end{equation}
at frequency $\nu=1$~GHz as a function of SNR diameter $D=2R_\mathrm{s}$.
In the sequel we shall use $\Sigma_\mathrm{R}$ for the brightness in the radio 
band and $\Sigma_\mathrm{X}$ for the X-ray
band, respectively.
Calculations were performed for the above three different ISM densities and 
for the three values of the SN explosion energy 
$E_\mathrm{sn}/(10^{51})~\mbox{erg}=0.25$, 1 and 3, respectively. 
It was assumed that this
energy interval covers the majority of the observed SNRs.
The experimental values corresponding to
37 Galactic shell SNR with known distances
(Case \& Bhattacharya 1998) and to SNRs in the Magellanic Clouds, M31
and M 33 (Berkhuijsen 1986) are also shown in Fig.6.
%
%------------------------------------------------------------------------fig.6-
\begin{figure}
\centering
\includegraphics[width=7.5cm]{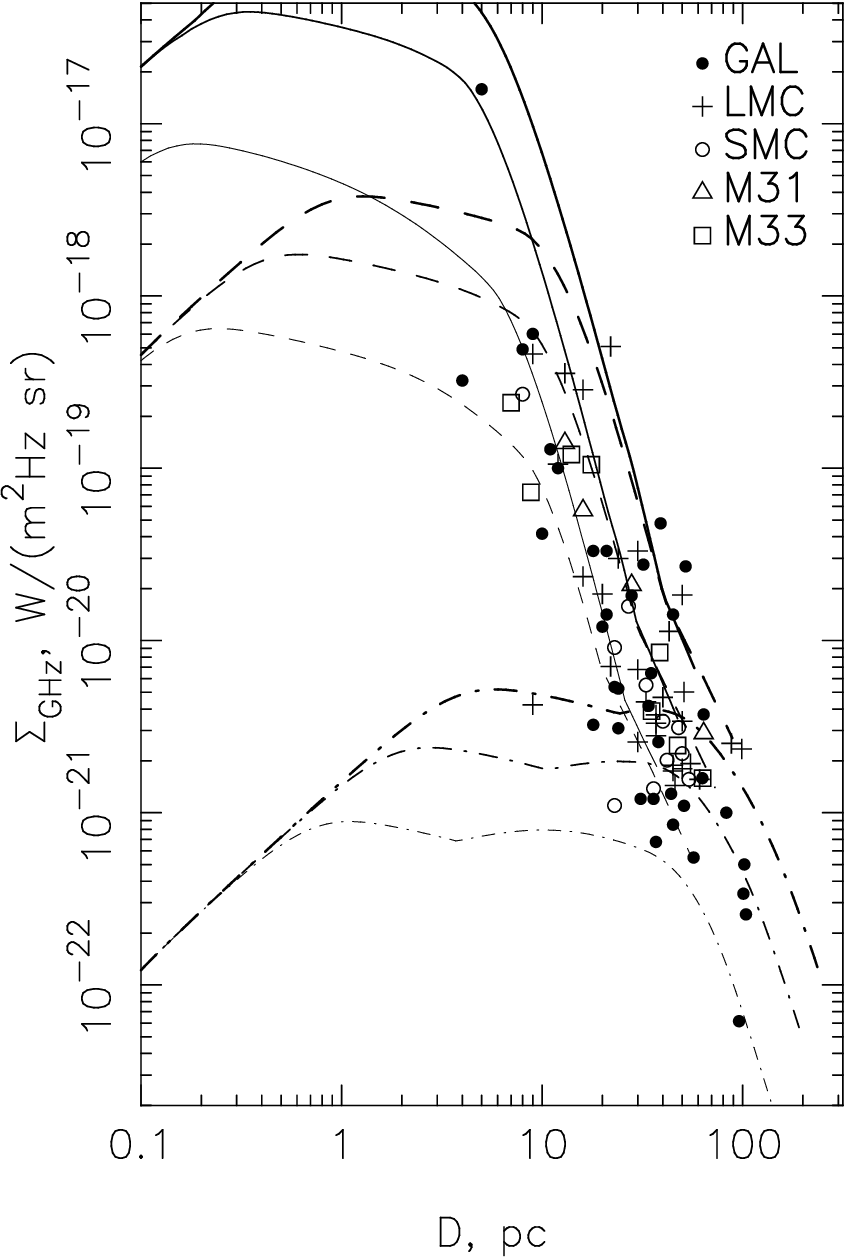}
\caption{Surface brightness-diameter diagram at frequency
  $\nu=1$~GHz. Different styles of lines correspond to the same cases
  as in Fig.4. Experimental data for the Galaxy
(Case \& Bhattacharya 1998) and for the  Magellanic Clouds, M~31 and M~33
(Berkhuijsen 1986) are shown.}
\end{figure}
%------------------------------------------------------------------------------
%

The relation $\Sigma_\mathrm{\nu}(D)$ is a direct consequence of the
$L_\mathrm{\nu}(D)$ dependence. During the initial part of the free expansion
phase according to Eq.(\ref{Lnuin}) the expected dependence is
\begin{equation}
\Sigma_\mathrm{R}\propto N_\mathrm{H}^{7/4} D,
\label{Sigmain}
\end{equation}
and explains very well the numerical $\Sigma_\mathrm{R}(D)$ behavior for
small $D$ ($D<0.2$~pc for $N_\mathrm{H}=3$~cm$^{-3}$ and $D<2$~pc for
$N_\mathrm{H}=0.003$~cm$^{-3}$) in Fig.6. In this region
$\Sigma_\mathrm{R}$ does not depend on the SN parameters $E_\mathrm{sn}$
and $M_\mathrm{ej}$, but does very sensitively depend on the ISM density.

During the subsequent part of the free
expansion phase the expected dependence is
\begin{equation}
\Sigma_\mathrm{R}\propto (E_\mathrm{sn}^2/M_\mathrm{ej})^{7/16}N_\mathrm{H}^{21/16}D^{-5/16},
\label{Sigmafe}
\end{equation}
cf. Eq.(\ref{Lnufe}).  During this period of SNR evolution the surface
brightness is very sensitive to the values of the explosion energy
$E_\mathrm{sn}$ and of the ambient gas number density $N_\mathrm{H}$, as can also be seen
from Fig.6.

In the subsequent Sedov phase, according to Eq. (\ref{Lnused}) we have 
\begin{equation}
\Sigma_\mathrm{R}\propto E_\mathrm{sn}^{7/4}D^{-17/4},
\label{Sigmased}
\end{equation}
independent of $N_\mathrm{H}$ and $M_\mathrm{ej}$. Since our magnetic
field strength 
$B_\mathrm{d}\propto V_\mathrm{s}$  in 
the
Sedov phase changes in a way similar to what was suggested by
Reynolds \& Chevalier (1981), the above $\Sigma_\mathrm{R}(D)$ dependence is
the same as they derived.

Note again that this dependence $\Sigma_\mathrm{R}\propto D^{-17/4}$ is valid only
for the electron spectrum $N_\mathrm{e}\propto p^{-2}$ which is created by the
unmodified shock. The actual shock during the Sedov phase, as can be seen
from Fig.1b, is significantly modified. The modification, characterized by
the deviation of the shock and subshock compression ratios $\sigma$ and
$\sigma_\mathrm{s}$ from the classical value 4, is larger for denser ISM. Therefore
the calculated dependence $\Sigma_\mathrm{R}(D)$ presented in Fig.6 is close to
$\Sigma_\mathrm{R}\propto D^{-17/4}$ in the case of a diluted ISM with number
density $N_\mathrm{H}=0.003$~cm$^{-3}$ for which the shock is only slightly
modified (see Fig.1b) and it becomes steeper with the increase of the ISM
density.

In the late Sedov epoch, when due to the shock deceleration the magnetic field 
amplification becomes irrelevant, and therefore the downstream SNR field
becomes constant, the $\Sigma_\mathrm{R}(D)$ dependence goes toward
\begin{equation}
\Sigma_\mathrm{R}\propto E_\mathrm{sn}N_\mathrm{H}^{3/4}D^{-2}.
\label{Sigmal}
\end{equation}

The fact that the observational data points lie in a relatively compact
region on the $\Sigma_\mathrm{R}-D$ diagram can be explained from a
theoretical point of view if we suggest that nearly all of the identified
SNRs with known distance are in the Sedov phase, or at least at the end
of the free expansion phase. The lack of a significant number of SNRs
detected in the early free expansion phase can be due to the small SNR
size and correspondingly small synchrotron flux expected in this stage,
as discussed before.

Even though the expected relation $\Sigma_\mathrm{R}\propto
E_\mathrm{sn}^{7/4}D^{-17/4}$ does not depend on the ISM gas density, it
is clear from Fig.6 that for small diameters $D<10$~pc only SNRs located
in a dense ISM with $N_\mathrm{H}>1$~cm$^{-3}$ are contained in the
sample, whereas for the largest diameters $D>30$~pc the dominant SNRs are
those which are in a tenuous ISM with $N_\mathrm{H}<0.01$~cm$^{-3}$. If in
addition we assume that SNRs in the early Sedov phase are predominantly
detected, then one would expect that the relation $D\propto
N_\mathrm{H}^{-1/3}$ is roughly valid, which leads to
$\Sigma_\mathrm{R}\propto N_\mathrm{H}^{17/12}$ which agrees very well
with the dependence determined from observations (Berkhuijsen 1986).

Note that type II and Ib SNe, whose number is expected to exceed
the number of type Ia SNe considerably, are characterized by essentially
different values of the mass of ejecta $M_\mathrm{ej}$ and of their 
initial velocity
distribution. Nevertheless, in the Sedov phase this difference becomes
irrelevant. Therefore we restrict our consideration here to type Ia SNe.
On the other hand, according to expressions (\ref{Vs(Rs)}) and 
(\ref{Lnufe}), in the free
expansion phase the expected radio luminosity of type II and
Ib SNe with typical ejecta masses between $M_\mathrm{ej}=5$ and $10
M_{\odot}$, is lower by a factor of two to three than that
of type Ia SNe considered here.

In Figs. 4 and 6 the brightest object is Cas~A, which is probably
a type Ib SN with a Wolf-Rayet star progenitor, or possibly a type II SN.
It is therefore not thought to genuinely belong to any of the
evolutionary tracks shown. However, it remains true that the enormous
radio luminosity and surface brightness of this young and rather small
object can only be explained by a high-density environment, as suggested
in these Figs.

\section{The $\Sigma-D$ relation in the X-ray range}
During the last several years a number of SNRs have been discovered as sources
of nonthermal X-rays (e.g. Petre et al. 2001). Studies of the spatial fine
structures of the nonthermal X-ray emission from SNRs with high angular
resolution instruments like Chandra (Hwang et al. 2002; Vink \& Laming
2003; Long et al. 2003; Bamba et al. 2003) are extremely interesting,
since the observed small scale structures yield decisive information about
the actual magnetic field strength in these remnants. The most important
conclusion is that in all cases the magnetic field strength derived from
these observations are considerably higher than expected for typical ISM
fields (Berezhko et al. 2003a; Berezhko \& V\"olk 2004).
%
%------------------------------------------------------------------------fig.7-
\begin{figure}
\centering
\includegraphics[width=7.5cm]{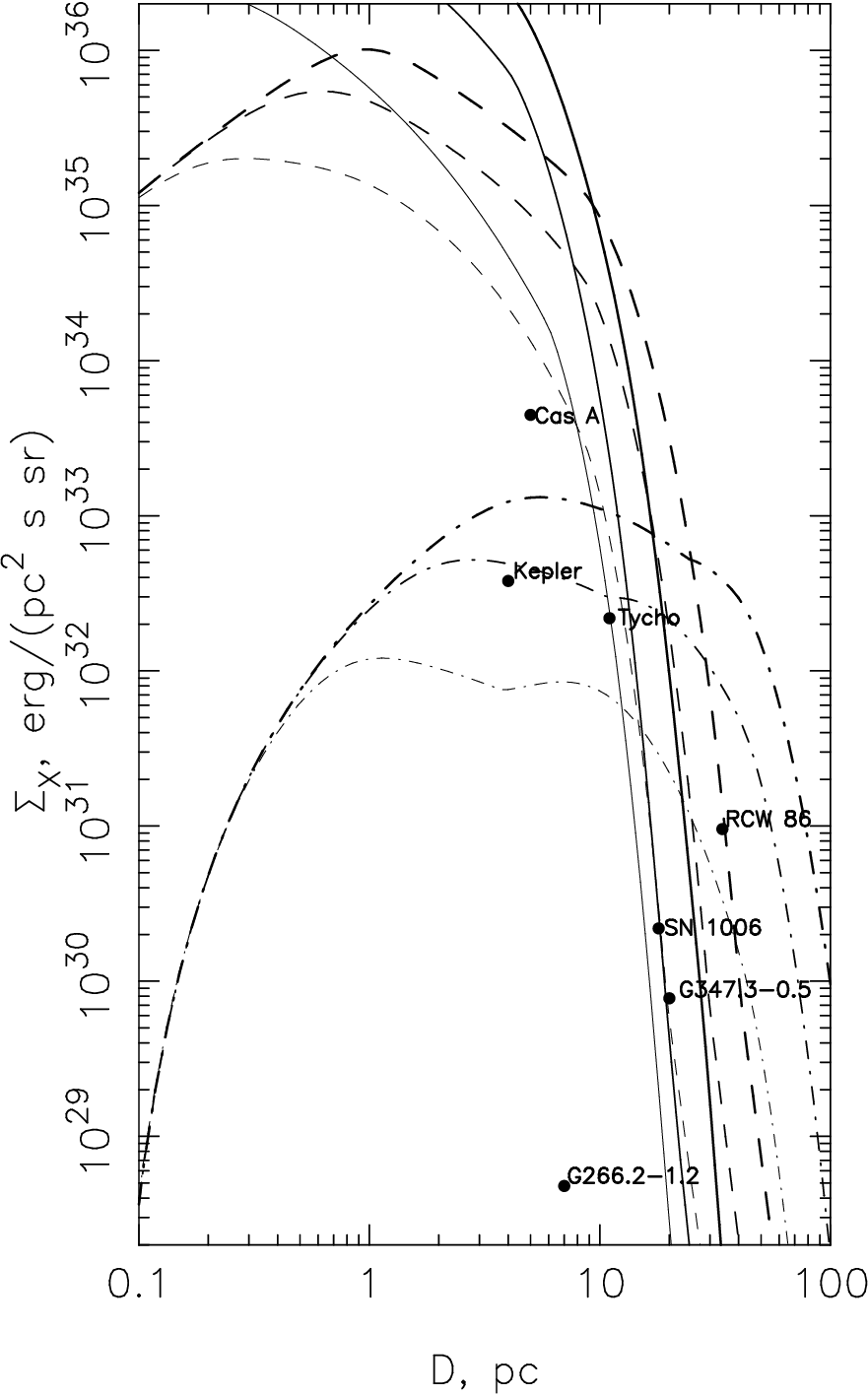}
\caption{The same as in Fig.6 but for the frequency
  $\nu=10^{18}$~Hz
(photon energy $\epsilon_\mathrm{\nu}\approx 4$~keV). Experimental values for 
seven Galactic SNRs (see text) are shown.}
\end{figure}
%------------------------------------------------------------------------------
%
Therefore it is worthwhile to study the synchrotron luminosity of SNRs 
in the X-ray band.

In Fig.7 we show the surface brightness $\Sigma_\mathrm{X}$ at frequency
$\nu=10^{18}$~Hz (photon energy $\epsilon_\mathrm{\nu}=4$~keV) as a function of
SNR diameter, for the same three values of ISM density and SN explosion
energy as considered before. The experimental brightness calculated from
the measured X-ray fluxes (and extrapolated to a given energy
$\epsilon_\mathrm{\nu}=4$~keV) for Cas~A, Tycho's SNR, RCW~86, Kepler and
SN~1006 is taken from Petre et al. (2001), for G347.3-0.5 from Slane et
al. (1999), and for G266.2-1.2 from Duncan \& Green (2000). A distance
$d=1$~kpc is taken for the last two objects.

In the free expansion phase the $\Sigma_\mathrm{X}(D)$ dependence is very similar
to that in the radio range because during this epoch the frequency is
lower than the cutoff frequency $\nu_\mathrm{max}$ (see Fig.3), and therefore
the X-ray luminosity is determined by the same factors as in the radio
band.

A dramatically different situation exists for larger diameters ($D>4$, 8
and 20~pc for $N_\mathrm{H}=1$, 0.3 and 0.003~cm$^{-3}$, respectively),
which correspond to the Sedov phase. Due to the decrease of the cutoff
frequency $\nu_\mathrm{max}$ the frequency $\nu=10^{18}$~Hz becomes larger
than $\nu_\mathrm{max}$ already at the beginning of this stage. This leads
to a very rapid decrease of the synchrotron luminosity $L_\mathrm{\nu}$
and of the brightness $\Sigma_\mathrm{X}$. For example, the brightness
$\Sigma_\mathrm{X}$ becomes as low as $10^{29}$~erg/(pc$^2$ s sr) for
$D=2$~pc in the case $E_\mathrm{sn}=10^{51}$~erg and
$N_\mathrm{H}=0.3$~cm$^{-3}$ that corresponds to a relatively young SNR
(see Fig.1a) of age $t\approx 2\times10^3$~yr. It is therefore not
surprising that the number of SNRs detected as sources of nonthermal
X-rays is so small, and that all of these sources are young.

Since the radio and X-ray synchrotron emissivities of SNRs are of the
same physical nature, it is interesting to study the relation to be 
expected
between them. For this purpose the radio brightness
$\Sigma_{1~\mbox{GHz}}$ is plotted in Fig.8 versus the X-ray SNR
brightness $\Sigma_\mathrm{X}$ for the same ranges of the explosion energy
$E_\mathrm{sn}$ and ISM number density $N_\mathrm{H}$ as before.

%------------------------------------------------------------------------fig.8-
\begin{figure}
\centering
\includegraphics[width=7.5cm]{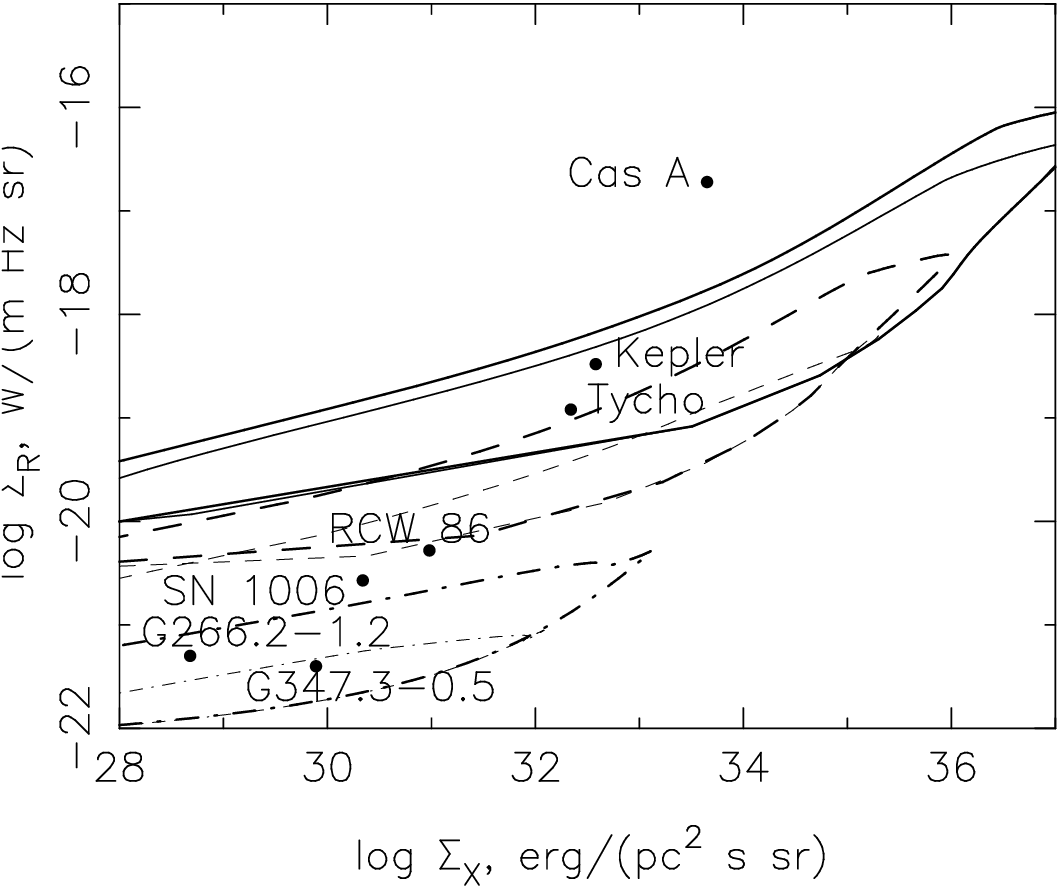}
\caption{Spectral surface brightness distribution at frequency $\nu=1$~GHz 
versus surface brightness
at $\nu=10^{18}$~Hz
(photon energy $\epsilon_\mathrm{\nu}\approx 4$~keV) for the same cases as in 
Fig.4, except that corresponding to $E_\mathrm{sn}=10^{51}$~erg.}
\end{figure}
%------------------------------------------------------------------------------
%

It is interesting to note that the allowed region in this
$\Sigma_\mathrm{R-}\Sigma_\mathrm{X}$ diagram is very limited, especially
if the ambient ISM density is known. This is because a relatively high
X-ray brightness is expected for only a limited range of the SNR sizes,
corresponding to the free expansion phase.  In Fig.8 all Galactic SNRs
except Cas~A are inside the expected region in the
$\Sigma_\mathrm{R}-\Sigma_\mathrm{X}$ diagram. A possible explanation is
again that Cas~A is not a type Ia SNR, but certainly a type Ib or a type
II SNR expanding into a nonuniform circumstellar medium strongly modified
by the progenitor star wind. What is even more essential in this case is
that the mass of the ejecta is substantially higher than in the case of a
type Ia SN. This implies a lower shock speed for the same explosion energy
than in the type Ia case. As was already pointed out, X-ray synchrotron
emission is very sensitive to the shock speed. Lowering the shock speed
will therefore shift the theoretical curve in Fig.8 mainly to the left,
toward lower values of $\Sigma_\mathrm{X}$. This could lead to bounding
the experimental Cas~A point by the curve that corresponds to
$N_\mathrm{H}=3$~cm$^{-3}$. This mean density appears to be an appropriate
value for Cas~A.

\section{Discussion and conclusions}
The nonlinear kinetic theory for CR production in SNRs has been used for
the first time in order to study the statistical properties of the
synchrotron emission of SNRs, in particular the $\Sigma-D$ relation.

Following the result of theoretical modeling of magnetic field
amplification near the shock due to CRs (Luceck \& Bell 2000), the
theoretical analysis of the spectral properties of the nonthermal emission
of individual SNRs (Berezhko et al. 2002, 2003b; V\"olk et al. 2002) which
predicted magnetic field strengths inside young SNRs much larger than
typical ISM values, and the experimental studies of the fine structure of
nonthermal X-ray emission of young SNRs (Hwang et al. 2002; Vink \& Laming
2003; Long et al.  2003; Bamba et al. 2003), we incorporate into our model
a time-dependent magnetic field, whose energy density is proportional to
the CR pressure. This effect influences the SNR synchrotron emission
drastically, making it much more intense, especially in the case of
relatively young SNRs.

Since the value of the magnetic field in the most active SNR phase for 
CR production, the early Sedov phase, is more than an order of
magnitude larger than the ISM field, the overall CR proton
power law spectrum, produced during the whole SNR evolution, extends
up to an energy $\epsilon_\mathrm{max}\approx 10^{15}$~eV, independent of
the ISM density. This is just the condition required to explain 
the Galactic CR spectrum due to SNRs up to the energy of the knee, which is
$3\times 10^{15}$~eV (Berezhko \& Ksenofontov 1999).

The maximum energy of CR electrons is about an order of magnitude lower
due to synchrotron losses. In addition, as a consequence of these losses,
the electron spectrum is much steeper than the proton spectrum at
energies $\epsilon_\mathrm{e}\gsim 100$~GeV.

Reynolds \& Keohane (1999) and Heindrich \& Reynolds (2001) estimated the
maximum electron energy for a number of galactic and extragalactic SNRs
based on the measured radio and X-ray spectra. Their values of
$\epsilon_\mathrm{max}^\mathrm{e}$ are within the range of from
$10\sqrt{10~\mu\mbox{G}/B_\mathrm{d}}$~TeV to
$200\sqrt{10~\mu\mbox{G}/B_\mathrm{d}}$~TeV, consistent with our
predictions (see Fig.2 and Fig.1a).

It was shown that, as a result of the progressively increasing overall
number of accelerated CRs, the radio synchrotron luminosity increases with
time in the free expansion phase, achieves its peak value at the very
beginning of the Sedov phase, and then again decreases with time due to
the decreasing strength of the effective magnetic field. The dependence of
the radio luminosity on the diameter, calculated for typical values of the
ISM density and of the SN explosion energy, covers the region of the
experimental points on the $L_\mathrm{\nu}-D$ diagram very well. The lack
of data for small linear sizes $D\ll 10$~pc is tentatively explained by a
low luminosity in the free expansion phase that makes it difficult to
recognize SNRs with small size and low luminosity.

In contrast to the radio range, the synchrotron luminosity in the X-ray
range, which has its peak value at the beginning of the Sedov phase as well, 
goes down much more rapidly with increasing $D$, due to the
decrease of the electron cutoff momentum. This determines the
corresponding value of the cutoff synchrotron frequency.

The theory predicts a dependence of radio SNR brightnesses on diameter $D$
which is close to $\Sigma_\mathrm{R}=A D^{-17/4}$ in the Sedov phase. All
the SNRs corresponding to the same explosion energy $E_\mathrm{sn}$ are
expected to align along the same line $\Sigma_\mathrm{R}=A D^{-17/4}$,
independent of ISM density. The amplitude of this dependence scales with
the explosion energy according to $A\propto E_\mathrm{sn}^{7/4}$.
Therefore the variation of $E_\mathrm{sn}$ within a factor of ten,
that has to be considered as an appropriate assumption for the actual
Galactic SNR population, makes it possible to cover almost the whole
region in the $\Sigma_\mathrm{R}-D$ diagram by the lines
$\Sigma_\mathrm{R}=A D^{-17/4}$. The spread in SNR luminosities at a given
SNR diameter $D$ is mainly due to the spread of the explosion energy
$E_\mathrm{sn}$. It therefore provides the theoretical basis for the
$\Sigma_\mathrm{R}-D$ diagram as an instrument for the distance
determination to SNRs.

The expected injection rate $\eta\sim 10^{-4}$ leads to significant SN
shock modification as a result of the CR backreaction during the Sedov
phase. Due to the strong nonlinear coupling at this stage the final
results are relatively insensitive to $\eta$. The synchrotron luminosity
is sensitive to $\eta$ only during the initial part of the free expansion
phase, when the shock is not modified, scaling roughly as
$L_\mathrm{\nu}\propto \eta^{7/4}$.

It is expected that SNRs in a dense ISM populate the part of the line
$\Sigma_\mathrm{R}=A D^{-17/4}$ which corresponds to low diameter values $D$,
whereas SNRs in a diluted ISM are expected to be on this line for
significantly larger $D$. This is an explanation for the empirical
correlation $\Sigma_\mathrm{R}\propto N_\mathrm{H}^{1.37}$ between the radio brightness
$\Sigma_\mathrm{R}$ and the ISM number density $N_\mathrm{H}$ (Berkhuijsen 1986). Since the
scale value $R_\mathrm{0}$ for the SNR size is proportional to $N_\mathrm{H}^{1/3}$, the
expected correlation is $\Sigma_\mathrm{R}\propto N_\mathrm{H}^{17/12}$. It agrees with the
experimental relation in a very satisfactory way.

\begin{acknowledgements}
This work has been supported in part by the Russian Foundation for Basic
Research (grant 03-02-16524). EGB acknowledges the hospitality of the
Max-Planck-Institut f\"ur Kernphysik where part of this work was carried
out.
\end{acknowledgements}

\appendix
\section{ }
We consider here the spatial dependence of the effective downstream
magnetic field $B_\mathrm{d}$ in the SNR interior.

Since the spatial scales of the amplified magnetic field are much smaller
than the shock radius $R_\mathrm{s}$, the field can be 
considered random and isotropic. Assuming it to be frozen into the thermal 
gas in 
the downstream region $r<R_\mathrm{s}$, one can write (e.g. Mestel 1965; Chevalier 
1974):
\begin{equation}
P_\mathrm{B}\rho^{-4/3} = const,
\end{equation}
\label{PBrho}
where $P_\mathrm{B}=B^2/(8\pi)$ is the magnetic pressure.
Therefore at any given radial distance $r$ we have
\begin{equation}
P_\mathrm{B}(r,t)\propto P_\mathrm{B2}(t_\mathrm{s})\rho(r,t)^{4/3},
\label{PB}
\end{equation}
where $t_\mathrm{s}$ is the instant of time when the gas element, at the present
time $t>t_\mathrm{s}$ located at radius $r$, had intersected the shock front,
$r(t_\mathrm{s})=R_\mathrm{s}(t_\mathrm{s})$. The postshock gas density $\rho_2$ is approximated by a
constant value $\rho_2=4\rho_0 $, where $\rho_0$ is the preshock density.

The profile of the downstream mass velocity $w(r,t)$ and density
$\rho(r,t)$ of the gas in the Sedov phase can be approximated by
$w=3rV_\mathrm{s}/(4R_\mathrm{s})$ and $\rho=4 \rho_0 (r/R_\mathrm{s}(t))^9$, neglecting here the
shock modification by the CR backreaction. The shock radius $R_\mathrm{s}$ and the
shock velocity $V_\mathrm{s}$ are given by $R_\mathrm{s}\propto t^{2/5}$ and $V_\mathrm{s}(t_\mathrm{s})\propto
t_\mathrm{s}^{-3/5}$, respectively. Therefore the equation of motion of a gas
element
\begin{equation}
dr/dt = w(r,t)
\end{equation}
\label{RDOT}   
integrates to
\begin{equation}
r/R_\mathrm{s}(t)=(t_\mathrm{s}/t)^{1/10},  
\end{equation}  
\label{RT} 
which gives $t_\mathrm{s}\propto r^{10}/t^3$.

Substituting $P_\mathrm{B2}(t_\mathrm{s})\propto V_\mathrm{s}^2(t_\mathrm{s})\propto t_\mathrm{s}^{-6/5}$ and
$t_\mathrm{s}(r,t)$ into Eq. (\ref{PB}), we find that the strong decrease of
$\rho^{2/3} \propto r^6$ into the interior, where fluid elements shocked
in the past are located, is compensated by the strong increase of $V_\mathrm{s}$ in
the same past, i.e. that $B_\mathrm{d}(r,t)=B_2(t)$. Thus $B_\mathrm{d}(r,t)$ is uniform in
this approximation.


\begin{thebibliography}{99}


\bibitem[2003]{bamba}
Bamba, A., Yamazaki, R., Ueno, M., Koyama, K.\ 2003, ApJ, 589, 827

\bibitem[1978]{bell78}
Bell, A.R. 1978, MNRAS, 182, 147

\bibitem[2001]{belll}
Bell, A. R. \& Lucek, S.G. 2001, MNRAS, 321, 433.

\bibitem[95]{benel}
Bennet, L. \& Ellison, D.C. 1995, JGR, 100, 3439

\bibitem[1988]{bk88}
Berezhko, E.G. \& Krymsky, G.F. 1988, Soviet Phys.-Uspekhi, 12, 155

\bibitem[1996]{ber96}
Berezhko, E.G. 1996 Astropart.\ Phys., 5, 367

\bibitem[1996]{byk96}
Berezhko, E.G., Elshin, V.K. \& Ksenofontov, L.T.1996 JETPh, 82, 1

\bibitem[1997]{bv97}
Berezhko, E.G. \& V\"olk, H.J. 1997, Astropart.\ Phys., 7, 183

\bibitem[1999]{berel}
Berezhko, E.G. \& Ellison, D.C. 1999, ApJ, 526, 385

\bibitem[1999]{bk99}
Berezhko, E.G. \& Ksenofontov, L.T. 1999, JETPh 89, 391

\bibitem[2002]{bkv02}
Berezhko, E.\ G., Ksenofontov, L.\ T., V\"olk, H.\ J.\ 2002, A\&A,
395, 943

\bibitem[2003a]{bkv03}
Berezhko, E.\ G., Ksenofontov, L.\ T., V\"olk, H.\ J.\ 2003a, A\&A,
412, L11

\bibitem[2003b]{bpv03}
Berezhko, E.\ G., P\"uhlhofer, G., V\"olk, H.\ J.\ 2003b, A\&A,
400, 971

\bibitem[2004]{bv04}
Berezhko, E.G. \& V\"olk, H.J. 2004, A\&A 419, L27

\bibitem[1990]{brz90}
Berezinskii, V.S., Bulanov, S.A., Dogel, V.A. et al. 1990,
 Astrophysics of cosmic rays, 
North-Holland: Publ.Comp.

\bibitem[1986]{bern}
Bernhuijsen, E.M. 1987, A\&A 166, 257

\bibitem[1978]{bo78}
Blandford, R.D. \& Ostriker, J.P. 1978, ApJ, 221, L29


\bibitem[1978]{bo87}  
Blandford, R.D. \& Eichler, D. 1987, Phys. Rept. 154, 1

\bibitem[1998]{case}
Case, G.L. \& Bhattacharya, D. 1998, ApJ 504, 761

\bibitem[1974]{chev74}
Chevalier, R.A. 1974, ApJ 188, 501

\bibitem[1982]{chev81}
Chevalier, R.A. 1982, ApJ, 258, 790

\bibitem[1984]{chevf}
Chevalier, R.A. \& Fransson, C. 1984, ApJ 420, 268

\bibitem[1983]{drury}
Drury, L'O.C. 1983, Rep. Progr. Phys. 46, 973

\bibitem[1986]{duric}
Duric, N. \& Seaquist, E.R. 1986, ApJ, 301, 308

\bibitem[2000]{duncan}
Duncan, A.R. \& Green, D.A. 2000, A\&A, 364, 732

\bibitem[1998]{dwach}
Dwarkadas, V.V. \& Chevalier, R.A. 1998, ApJ, 497, 807

\bibitem[1983]{fedor}
Fedorenko, V.N. 1983, in IAU Symp. 101, Supernova Remnants and their
X-ray Emission, ed. J.Danziger \& P.Gorenstein (Dordrecht: Reidel), 183

\bibitem[1984]{green}
Green, D.A. 1984, MNRAS 209, 449

\bibitem[1994]{huang94}
Huang, Z.P., Thun, T.X., Chevalier, R.A. et al. 1994, ApJ 424, 114
 
\bibitem[2002]{hwang}
Hwang, U., Decourshelle, A., Holt, S.S. \& Petre, R. 2002, ApJ, 581, 1101

\bibitem[2001]{hr01}
Heindrich, S.P. \& Reynolds, S. P. 2001, ApJ 559, 903

\bibitem[1981]{Jones81}
Jones, E.M., Smith, B.W. \& Straka, W.C. 1981, ApJ, 249, 185

\bibitem[1995]{koyama}
Koyama, K., Petre, R., Gotthelf, E.V. et al. 1995, Nature, 378, 255

\bibitem[2004]{ksen}
Ksenofontov, L.T., Berezhko, E.G. \& V\"olk, H.J. 2004, submitted to A\&A

\bibitem[2003]{long}
Long, K.S., Reynolds, S.P., Raymond, J.C., et al.\ 2003, ApJ, 586, 1162

\bibitem[200]{lucb00}
Lucek, S.G. \& Bell, A.R. 2000, MNRAS, 314, 65

\bibitem[1998]{malk}
Malkov, M.A. 1998, Phys. Rev. E, 58, 4911

\bibitem[1995]{malkv}
Malkov, M.A. \& V\"olk, H.J. 1995, A\&A, 300, 605

\bibitem[1982]{mck82}
McKenzie, J.F. \& V\"olk, H.J. 1982, A\&A, 116, 191

\bibitem[1965]{mes65}
Mestel, L. 1965, Quart. J. Roy. Astr. Soc., 6, 161 and 265

\bibitem[2001]{petre}
Petre, R., Hwang, U. \& Allen, G.E. 2001, Adv. Space Sci. 27, 647

\bibitem[2003]{pz}
Ptuskin, V.S. \& Zirakashvili, V.N. 2003, A\&A, 403, 1

\bibitem[1981]{reynch}
Reynolds, S.P. \& Chevalier, R.A. 1981, ApJ, 245, 912

\bibitem[1992]{reyel}
Reynolds, S.P. \& Ellison, D.C. 1992, ApJ, 399, L75

\bibitem[1999]{rk99}
Reynolds, S.P. \& Keohane, J.W. 1999, ApJ 525, 368

\bibitem[1992]{scholer}
Scholer, M., Trattner, K.J. \& Kucharek, H. 1992, ApJ, 395, 675

\bibitem[1999]{slane}
Slane, P., Gaensler, B.M., Dame, T.M. et al. 1999, ApJ 525, 357

\bibitem[1994]{tamm}
Tammann, G.A., L\"ofler, W. \& Schr\"oder, A. 1994, ApJ, 92, 487.

\bibitem[1994]{trat}
Trattner, K.J. \& Scholer, M. 1994, JGR, 99, 6637

\bibitem[2003]{vl03}
Vink, J. \& Laming, J.M. 2003, ApJ, 548, 758

\bibitem[2002]{vbkr}
V\"olk, H.J., Berezhko, E.G., Ksenofontov, L.T. \& Rowell, G.P. 2002,
A\&A 396, 971

\bibitem[2003]{vbk03}
V\"olk, H.J., Berezhko, E.G. \& Ksenofontov, L.T. 2003, A\&A, 409,
563

\end{thebibliography}
\end{document}